\documentclass[12pt,preprint]{aastex}

\newcommand{\muuu}{$\Sigma_{160 \mu {\rm m}}$}
\newcommand{\ropt}{R$_{25}$}
\newcommand{\halpha}{H$\alpha$}

\shorttitle{The {\it Herschel} KINGFISH Project}
\shortauthors{Kennicutt et al.}

\begin{document}


\title{KINGFISH -- Key Insights on Nearby Galaxies: A Far-Infrared Survey 
with {\it Herschel}: Survey Description and Image Atlas\footnote{{\it Herschel} is an ESA space observatory with science instruments provided by European-led Principal Investigator consortia and with important participation from NASA.}} 



\author{
R.~C. Kennicutt\altaffilmark{1},
D. Calzetti\altaffilmark{2},
G. Aniano\altaffilmark{3},
P. Appleton\altaffilmark{4},
L. Armus\altaffilmark{5},
P. Beir\~{a}o\altaffilmark{5},
A.~D. Bolatto\altaffilmark{6},
B. Brandl\altaffilmark{7},
A. Crocker\altaffilmark{2},
K. Croxall\altaffilmark{8},
D.~A. Dale\altaffilmark{9},
J. Dononvan Meyer\altaffilmark{10},
B.~T. Draine\altaffilmark{3},
C.~W. Engelbracht\altaffilmark{11},
M. Galametz\altaffilmark{1},
K.~D. Gordon\altaffilmark{12},
B. Groves\altaffilmark{7,16},
C.-N. Hao\altaffilmark{13},
G. Helou\altaffilmark{4},
J. Hinz\altaffilmark{11},
L.~K. Hunt\altaffilmark{14},
B. Johnson\altaffilmark{15},
J. Koda\altaffilmark{10},
O. Krause\altaffilmark{16}, 
A.~K. Leroy\altaffilmark{17},
Y. Li\altaffilmark{2},
S. Meidt\altaffilmark{16},
E. Montiel\altaffilmark{11}, 
E.~J. Murphy\altaffilmark{18}, 
N. Rahman\altaffilmark{6},
H.-W. Rix\altaffilmark{16},
H. Roussel\altaffilmark{15},
K. Sandstrom\altaffilmark{16},
M. Sauvage\altaffilmark{19},
E. Schinnerer\altaffilmark{16},
R. Skibba\altaffilmark{11},
J.~D.T. Smith\altaffilmark{8},
S. Srinivasan\altaffilmark{15},
L. Vigroux\altaffilmark{15},
F. Walter\altaffilmark{16},
C.~D. Wilson\altaffilmark{20},
M. Wolfire\altaffilmark{6},
S. Zibetti\altaffilmark{21}
}

\altaffiltext{1}{Institute of Astronomy, University of Cambridge, Madingley Road,
Cambridge CB3 0HA, UK}

\altaffiltext{2}{Department of Astronomy, University of Massachusetts, Amherst, MA 01003, USA}

\altaffiltext{3}{Department of Astrophysical Sciences, Princeton University,
Princeton, NJ 08544, USA}

\altaffiltext{4}{NASA Herschel Science Center, IPAC, California Institute of
Technology, Pasadena, CA 91125, USA}

\altaffiltext{5}{Spitzer Science Center, California Institute of Technology, MC
314-6, Pasadena, CA 91125, USA}

\altaffiltext{6}{Department of Astronomy, University of Maryland, College Park, MD 20742, USA}

\altaffiltext{7}{Leiden Observatory, Leiden University, P.O. Box 9513, 2300 RA
Leiden, The Netherlands}

\altaffiltext{8}{Department of Physics and Astronomy, University of
Toledo, Toledo, OH 43606, USA}

\altaffiltext{9}{Department of Physics \& Astronomy, University of Wyoming,
Laramie, WY 82071, USA}

\altaffiltext{10}{Department of Physics and Astronomy, SUNY Stony Brook, Stony
Brook, NY 11794-3800, USA}

\altaffiltext{11}{Steward Observatory, University of Arizona, Tucson, AZ 85721, USA}

\altaffiltext{12}{Space Telescope Science Institute, 3700 San Martin Drive, Baltimore,
MD  21218, USA}

\altaffiltext{13}{Tianjin Astrophysics Center, Tianjin Normal University, Tianjin
300387, China}

\altaffiltext{14}{INAF - Osservatorio Astrofisico di Arcetri, Largo E. Fermi 5,
50125 Firenze, Italy}

\altaffiltext{15}{Institut d'Astrophysique de Paris, UMR7095 CNRS, Universit\'e
Pierre \& Marie Curie, 98 bis Boulevard Arago, 75014 Paris, France}

\altaffiltext{16}{Max-Planck-Institut f\"ur Astronomie, K\"onigstuhl 17, D-69117 Heidelberg, Germany} 

\altaffiltext{17}{National Radio Astronomy Observatory, 520 Edgemont Road, Charlottesville, VA 22903, USA}

\altaffiltext{18}{Observatories of the Carnegie Institution for Science, 813 Santa Barbara Street, Pasadena, CA 91101 USA}

\altaffiltext{19}{CEA/DSM/DAPNIA/Service d'Astrophysique, UMR AIM, CE Saclay, 91191 Gif sur Yvette Cedex}

\altaffiltext{20}{Department of Physics \& Astronomy, McMaster University,
Hamilton, Ontario L8S 4M1, Canada}

\altaffiltext{21}{DARK Cosmology Centre, Niels Bohr Institute, University of Copenhagen, Juliane Maries Vej 30, DK-2100 Copenhagen, Denmark}


\begin{abstract}

The KINGFISH project (Key Insights on Nearby Galaxies: a Far-Infrared 
Survey with {\it Herschel}) is an imaging and spectroscopic survey of 61 
nearby ($d < 30$ Mpc) galaxies, chosen to cover a wide range of 
galaxy properties and local interstellar medium (ISM) environments 
found in the nearby 
Universe.   Its broad goals are to characterize the ISM of present-day 
galaxies, the heating and cooling of their gaseous and dust 
components, and to better understand the physical processes linking 
star formation and the ISM.  KINGFISH is a direct descendant of the {\it Spitzer} 
Infrared Nearby Galaxies Survey (SINGS), which produced complete 
Spitzer imaging and spectroscopic mapping and a comprehensive set
of multi-wavelength ancillary observations for the sample.
The {\it Herschel} imaging consists of complete maps for the galaxies 
at 70, 100, 160, 250, 350, and 500\,$\mu$m.  The spectal line  
imaging of the principal atomic ISM cooling
lines ([OI]63\,$\mu$m, [OIII]88\,$\mu$m, [NII]122,205\,$\mu$m, and 
[CII]158\,$\mu$m) covers the subregions in the centers and disks that already
have been mapped in the mid-infrared with Spitzer.
The KINGFISH and SINGS multi-wavelength datasets combined provide panchromatic
mapping of the galaxies sufficient to resolve individual star-forming regions,
and tracing the important heating and cooling channels of 
the ISM, across a wide range of local extragalactic ISM environments.  
This paper summarizes the scientific strategy for KINGFISH, the
properties of the galaxy sample, the observing strategy, and
data processing and products.  It also presents a combined {\it Spitzer}
and {\it Herschel} image atlas for the KINGFISH galaxies, covering
the wavelength range 3.6 -- 500\,\micron.  All imaging and
spectroscopy data products will be released to the {\it Herschel}
user generated product archives.

\end{abstract}





\section{INTRODUCTION}

Observations of star formation and the interstellar medium (ISM) in nearby 
galaxies form a vital bridge between in-depth studies of individual interstellar
clouds and star-forming regions in the Galaxy and the globally 
integrated measurements of distant galaxies.  Many of the physical
processes which are key to triggering and regulating star formation
are manifested on kiloparsec or sub-kiloparsec scales.  These 
include the formation and cooling of the atomic gas, the formation
of molecular gas and gravitationally bound clouds, the fragmentation
of clouds to form bound cores, stars, and star clusters, and the return
of radiant and mechanical energy from those stars into the ISM.  
Until recently this complex interplay of physical processes could
only be probed in depth in the Galaxy and close neighbors such as
the Magellanic Clouds.  The advent of groundbased aperture synthesis
arrays at submillimeter to centimeter wavelengths and of sensitive
space telescopes in the ultraviolet to submillimeter range have
made it possible to extend such work to galaxies in the local
Universe, and thus to a range of galactic and interstellar environments
which was inaccessible only a few years ago.

These broad aims formed the basis of the {\it Spitzer} Infrared Nearby 
Galaxies Survey (SINGS; Kennicutt et al. 2003), one of the six
original {\it Spitzer} Legacy Science projects.  The primary goal of SINGS
was to fully characterize the infrared emission of galaxies and their 
principal infrared-emitting components, using imaging of 75 nearby
galaxies at 3.6--160\,$\mu$m along with a suite of spectroscopic mapping 
of the galaxy centers and selected sub-regions in the 5--40\,$\mu$m range.  
The galaxies
and the spectroscopic targets were selected to span broad ranges of 
physical properties, star formation properties, and local interstellar
environments.  The {\it Spitzer} observations of the SINGS sample were
supported by an extensive campaign of ancillary observations extending from X-rays
to the radio (Kennicutt et al. 2003), and have led to a number
of follow-up surveys (many led outside of the original SINGS
collaboration), including mapping of the ultraviolet continuum
emission (Gil de Paz et al. 2007), radio continuum emission
(Braun et al. 2007), velocity-resolved Fabry-Perot mapping of
\halpha\ emission (Daigle et al. 2006, Dicaire et al. 2008),
atomic and molecular gas (Walter et al. 2008; Leroy et al. 2009),
and X-ray emission (Jenkins et al. 2011), 
and optical spectroscopy (Moustakas et al. 2010).  
As the result of this follow-up work the SINGS sample offers the
most comprehensive multi-wavelength dataset for any local galaxy
sample of its size ever assembled.  Some of the main scientific
products of the survey have included comprehensive multi-wavelength
spectral energy distribution (SED) atlas (Dale et al. 2007), 
spectral atlases (Dale et al. 2009;
Moustakas et al. 2010); studies of the dust contents and polycyclic
aromatic hydrocarbon (PAH)
emission (Engelbracht et al. 2006; Draine et al. 2007; 
Prescott et al. 2007; Smith et al. 2007; Bendo et al. 2008;
Mu\~{n}oz-Mateos et al. 2009a, b) and the molecular gas emission 
(Roussel et al. 2007); studies of
the form and physical origins of the spatially-resolved radio-infrared
correlation (Murphy et al. 2006a, b; 2008); testing and
calibration of dust-corrected measurements of the star formation
rate (e.g., Calzetti et al. 2005, 2007, 2010; 
P\'{e}rez-Gonz\'{a}lez et al. 2006; Kennicutt et al. 2009, Li et al. 2010);
studies of the form of the star formation rate (SFR) vs
gas density law (e.g., Kennicutt et al. 2007; Bigiel et al. 2008; 
Leroy et al. 2008; Liu et al. 2011),
along with numerous studies of individual galaxies.  Most of
these investigations exploited the powerful synergies between
the infrared observations of the interstellar dust and/or gas
emission with coordinated measurements at other wavelengths.

These investigations are now reaching their culmination with 
the advent of the {\it Herschel Space Observatory} (Pilbratt et al. 2010).
For observations of nearby star-forming galaxies {\it Herschel's} 
key capabilities are its spatial resolution for far-infrared (FIR) imaging,
its extended wavelength coverage into the submillimeter,  and its spectroscopic
power for mapping the primary cooling transitions of the gaseous ISM.  Although 
{\it Spitzer} provided maps of the mid-infrared dust emission 
(5.8--24\,$\mu$m) with angular resolutions of 6\arcsec\ or better,
its beam sizes at 70\,$\mu$m and 160\,$\mu$m, where the bulk of
the dust emission is radiated, degrade to $\sim$18\arcsec\ and 40\arcsec\ FWHM,
respectively, corresponding to typical linear dimensions of 
1--5 kpc in the SINGS galaxies.  Thanks to a telescope aperture that is 
four times larger than {\it Spitzer}, the {\it Herschel} 
cameras image the peak of the thermal dust emission with
spatial resolutions of a few to several arcseconds, 
comparable to that of the {\it Spitzer} 24\,$\mu$m maps, and sufficient
to resolve the emission of dust-emitting clouds from the diffuse
disk background emission in most systems.    

The spectral region between 50\,$\mu$m and 250\,$\mu$m includes some of the
primary atomic cooling lines for the neutral and ionized
phases of the ISM, including [OI]63\micron, [OIII]88\micron, 
[NII]122,205\micron, and [CII]158\micron.  
These lines variously probe the photodissociation regions 
(PDRs) in the interfaces between molecular, atomic, and
ionized gas phases in star-forming regions, the emission
of the ionized gas itself in HII regions, and the diffuse
atomic gas.  When combined with spectral diagnostics from
the mid-infrared and visible these lines provide powerful
probes of the physical conditions and radiation fields
in the ISM, the heating of gas and dust, and potentially
robust empirical tracers of star-forming galaxies at 
high redshift.   

These unique scientific capabilities of {\it Herschel} can only
be fully exploited, however, if 
its imaging and spectroscopy are mated to a comprehensive  
set of observations at other wavelengths.  As such the SINGS
project provides an optimal foundational dataset for such
a study, and a follow-up {\it Herschel} survey, KINGFISH 
(Key Insights on Nearby Galaxies: a Far-Infrared Survey with {\it Herschel})
was proposed and approved as an Open-Time Key Programme.
With 536.6 hours of observing time it is the third largest
scientific programme being carried out with the {\it Herschel Space
Observatory}.

The KINGFISH project is 
an imaging and spectroscopic survey of 61 nearby ($d < 30$ Mpc)
galaxies (including 57 galaxies from the SINGS project),
chosen to cover the full range of integrated properties and
local ISM environments found in the nearby Universe.
The broad goals of KINGFISH are to characterize the ISM of 
present-day galaxies, the heating and cooling of
their gaseous and dust components, and to better understand
the physical processes linking star formation to the ISM.

The main aims of this paper are to provide basic information
on the core science program, sample selection and properties,
observing strategy, data properties and processing, and data
products, as a foundation for the numerous science exploitation
papers which will follow.  The remainder of this paper is
organized as follows.  In \S 2 we briefly summarize the 
scientific objectives for the project and demonstrate how
these translated into the experimental design for the {\it Herschel}
observations.  In \S 3 we describe the KINGFISH sample and observing strategy,
as well as the {\it Herschel} observations themselves, and (briefly)
the rich set of ancillary data available for the galaxies.
This section includes an updated tabulation of the distances
and physical properties of the galaxies, so future papers 
can be consistent in what is assumed.  In \S 4 we describe
the processing of the {\it Herschel} observations and the 
data products that we plan to deliver from the survey.
We also present examples
of early KINGFISH observations to illustrate the capabilities
of performance of the {\it Herschel} instruments for this application.
Other examples of early KINGFISH observations can be found
in the Science Demonstration Phase papers (Beirao et al. 2010,
Engelbracht et al. 2010, Sandstrom et al. 2010) and in Walter
et al. (2011), Murphy et al. (2011), and Skibba et al. (2011).

\section{SCIENTIFIC OBJECTIVES}

The KINGFISH science strategy is built around three principal
scientific objectives:  1)  A comprehensive study of the
dust-obscured component of star formation in galaxies, and
the relation between star formation and dust heating;
2)  A complete inventory of cold dust and its relation to
other dust components in the ISM;  3)  Spatially resolved
studies of the heating and cooling of the ISM, as traced
by atomic cooling lines and the dust.  The first objective
benefits foremost from {\it Herschel's} excellent imaging
resolution, the second from its long-wavelength imaging 
capabilities, and the third from its unprecedented spectroscopic
capabilities.  Each of these aims
is described in more detail in this section.

\subsection{Linking Star Formation to the Interstellar Medium in Galaxies} 

Approximately half of the bolometric luminosity of the Universe
is channeled through the far-infrared (FIR) emission of galaxies
(Lagache et al. 2005),
and the IR thus carries information on the full range of heating 
stellar populations, as well as on the structure and physical 
conditions of the absorbing dust itself. Dissecting this information
in practice is hampered by the highly clumped and variable
structure of the stars and dust, and the presence of multiple
dust components. 

The {\it Infrared Astronomical Satellite}
({\it IRAS}) established some basic global trends in the IR luminosities
and colors of galaxies as functions of type and mass, and revealed 
a separate class of starburst and AGN-driven IR-luminous galaxies
(e.g., Soifer et al. 1987).  The {\it Infrared Space Observatory (ISO)} 
and {\it Spitzer} Space Telescope missions brought the next breakthrough, by 
spatially resolving nearby galaxies in the mid-infrared, and by mapping 
separately the emission of individual dust grain components, each 
with its own stellar heating population (e.g., Helou et al. 2000; 
Draine et al. 2007 and references therein; Soifer et al. 2008).  These include
(1) warm sources associated with gas and dust clouds (HII regions)
surrounding young ($<$10 Myr old) star-forming regions 
(especially prominent in the {\it ISO} 15\micron\ and {\it Spitzer} 24\micron\ bands);  
(2)  a more diffuse, extended, and cooler dust component heated
by stars with a range of ages, which dominates the far-infrared emission
(the {\it IRAS} cirrus component); and (3)  mid-infrared band emission
from large PAH molecules, transiently heated by single UV and optical
photons in PDRs surrounding young star clusters and 
by the general interstellar radiation field.

These {\it Spitzer} and {\it ISO} observations have been especially
successful in establishing the physical connections between
the heating of the infrared-emitting dust and young stars, 
but they cannot, because of insufficient angular resolution, separate
the various emitting regions and components of most galaxies in the FIR, where
the bulk ($\sim$90\%) of the dust emission occurs.   The beam
sizes of the {\it Spitzer} MIPS (Multiband Imaging Photometer for {\it Spitzer}) 
imager, for example, are $\sim$18\arcsec\
and 40\arcsec\ FWHM at 70\,$\mu$m and 160\,$\mu$m, respectively,
corresponding to linear dimensions of 0.9 and 2.0 kpc for
a galaxy at a distance of 10 Mpc, similar to the median distance of
the galaxies in the SINGS and KINGFISH samples.   The larger aperture of
{\it Herschel} allows these wavelengths to be imaged with nearly
a fourfold improvement of spatial resolution, with point spread functions of
$\sim$5\farcs5 and 12\arcsec\ FWHM, respectively, at the
same wavelengths.  

The dramatic improvement in structural information with {\it Herschel}
is illustrated in Figure 1, which compares {\it Spitzer} MIPS
images of the SINGS/KINGFISH 
galaxy NGC\,628 (M74) at 70 and 160\,$\mu$m with {\it Herschel} PACS 
(Photodetector Array Camera and Spectrometer; Poglitsch et al. 2010) images at
the same wavelengths.  For galaxies at distances of $<$30 Mpc
(the limit for the SINGS and KINGFISH samples), this difference
in resolving power is critical, because it makes it possible to
separate star-forming and giant HII regions from quiescent regions, and resolve 
the nuclear, circumnuclear, and more extended disk emission.  This resolution
is also well matched to the physical scales over which cloud formation is 
triggered, and over which dust reprocesses the light of young stars
(e.g., Lawton et al. 2010).
The combination of {\it Herschel} FIR and submillimeter 
images with the SINGS images at shorter wavelengths provides spatially-resolved
SED maps extending from the UV to the FIR.  
%Figure 1 also shows 
%that despite the superior spatial resolution of PACS the {\it Spitzer} MIPS
%images are more sensitive to extended faint diffuse dust emission,
%owing to the much lower thermal backgrounds with MIPS.

\subsubsection{Understanding and Modelling Dust Heating and Emission in Galaxies}

The FIR and submillimeter SED maps from {\it Herschel} also 
make it possible to break some of the degeneracies that plague
the interpretation of current observations of dust in nearby galaxies.
The dust emission and temperature distribution is determined
by several factors including the local radiation field intensity,
dust opacity, grain size distribution, and composition.
{\it ISO} and {\it Spitzer} enabled a major advance by providing measurements 
to treat successfully the heating of the PAH grains and larger FIR-emitting grains
separately.  The limited wavelength coverage of those data, however,
make it difficult to separate the distributions of 
dust temperatures from the grain emissivity functions.  

{\it Herschel's} breakthrough capability for this problem is the deep 
submillimeter imaging capability offered by the SPIRE instrument
(Spectral and Photometric Imaging Receiver; Griffin et al. 2010).
SPIRE provides confusion-limited images at 250, 350, and 500\,$\mu$m,
sufficient to detect cool dust and constrain the Rayleigh-Jeans
region of the main dust emission components.  The FIR and submillimeter
SEDs can be fitted with
dust heating models (e.g., Draine \& Li 2007; Draine et al. 2007)
to probe dust emission at all ranges of temperatures (from warm
dust at T$\sim$100~K to cooler dust down to  T$\sim$15--20~K), to
test for changes in the wavelength-dependent emissivities with changing
metallicity, molecular/atomic gas fractions, and
local radiation field environments.  These results in turn will constrain
the composition and survival properties of the grains, to supplement studies 
of the Galactic ISM.  

The effectiveness of the broad wavelength coverage offered
by the combination of {\it Spitzer} and {\it Herschel} (PACS $+$ SPIRE)
mapping is illustrated in Figure 2, which shows the integrated
(full-galaxy) SED of the KINGFISH galaxy NGC\,337, with 
a Draine \& Li (2007) dust emission model superimposed.

The spatial resolution of the {\it Herschel} FIR maps will also 
make it possible to disentangle the respective roles of different
stellar populations in heating the dust.  
Different age stellar populations have distinct spatial
distributions within galaxies, and comparing the surface brightness 
and color distributions of the IR emission over a wide baseline
in wavelengths with those of different stellar populations (readily
traced by SINGS UV, broadband visible, and H$\alpha$ imaging)
will allow us to identify the dominant heating populations as a
function of location and wavelength, and to constrain
the role of other possible sources such as cosmic-ray heating
(e.g., Hinz et al. 2004, 2006).  The spatially-resolved dust SEDs in galaxies
are determined by a combination of factors, including the intensity
and spectral shape of the local radiation field, dust opacity,
and the heating geometry.  Separating these factors requires a sample
of galaxies with a wide range of SFRs, distributions of star formation,
and dust environments, and the KINGFISH sample was specifically
designed to offer this wide range of environments (\S3). 

The high--angular resolution {\it Herschel} data will also address another 
problem raised by {\it ISO} and {\it Spitzer} observations, namely the relation of
the mid-infrared PAH band emission at 3--18\,\micron\ to the other dust components,
and the stellar populations which are mainly responsible for powering 
this emission.  {\it Spitzer} and {\it ISO}
studies revealed at least two important sources
of PAH emission in star-forming galaxies:  emission from PDRs
surrounding star-forming regions, and more extended diffuse
PAH emission driven by the general interstellar radiation field, in relative
proportions which can vary significantly within and between galaxies 
(e.g., Helou et al. 2004).  Studies have
shown that the PAH emission correlates with both the cold (T$\sim 20$~K)
dust heated by the general, stellar population
(Haas et al. 2002, Bendo et al. 2008) on the one hand, and with the number 
of ionizing photons from the young stellar population on the other hand
(e.g., Roussel et al. 2001, F\"{o}rster Schreiber et al. 2004).
Reconciling these apparently discordant results is handicapped by the
limited angular resolution and sensitivity of the available maps of the 
diffuse cool dust component.  {\it Herschel} now is making it possible to
obtain maps of the cool dust component with spatial resolutions that
are more comparable to that of the 8\,$\mu$m PAH band emission, so
the stellar populations heating the respective dust components can
be compared directly.  Understanding the relationships between the PAH
emission, large-grain emission, and star formation is also important
because the PAHs are believed to be major contributors to the heating
of the interstellar gas, especially in PDR regions (\S2.3.1).

\subsubsection{Robust Multi-Wavelength Star Formation Rates and the Schmidt Law}  

The detection of large populations of dusty star-forming galaxies
at high redshift by {\it ISO}, {\it Spitzer}, and now by {\it Herschel}
has underscored the need for reliable calibrations of
IR-based and composite UV$+$IR or visible$+$IR based SFR diagnostics, 
and a physical
understanding of the basis and limitations of these tools
(Kennicutt 1998).  The SINGS project has confirmed 
the close association of the warm 24\,\micron\ dust component
with the youngest star-forming population in HII regions
(Calzetti et al. 2007), which has enabled the 
derivation of reliable, extinction-corrected SFR indicators 
(Calzetti et al. 2007, 2010; Rieke et al. 2009; Kennicutt et al. 2009).  
However the physical basis and applicability of such calibrations to 
high-redshift galaxies, which often exhibit vastly higher SFRs and
different interstellar environments than found locally, is unclear,
because the 24\,\micron\ emission comprises typically less than 10\%  
of the total dust emission of galaxies, and the spatial resolution
of {\it Spitzer} at longer wavelengths does not allow for the clean
resolution of individual star-forming complexes in most galaxies.

The sub-kpc scale FIR maps provided by the 
{\it Herschel} PACS instrument will directly elucidate the relation between
the 24\,$\mu$m emission and the FIR emission at the peak
of the bolometric dust emission, in the 70--170\,\micron\ range,
for individual star-forming regions.  This will make it possible
to calibrate directly SFR diagnostics based on composite 
UV+IR and H$\alpha$+IR tracers, for spatially resolved 
applications within galaxies, and for 
global studies of more distant galaxies.  The same data 
will quantify the random and systematic errors in SFRs based
on IR measurements alone (due to starlight not absorbed by
dust).  The resulting empirically-calibrated SFR recipes can
also be compared to those expected from fitting population synthesis
and dust models to the observed SED maps, to test the reliability of these 
methods in environments that differ substantially from those found in 
local galaxies (e.g., see Murphy et al. 2011).

The resolved attenuation-corrected {\it Herschel} SFR maps span 
a considerably wider dynamic range than is currently available 
from integrated SFR measurements (e.g., Kennicutt et al. 2009),
extending in particular to low surface brightness  
regimes where H$\alpha$ no longer necessarily provides a statistically reliable
local SFR measure (e.g., Cervi\~{n}o et al. 2003; Salim et al. 2007.
Meurer et al. 2009;
Lee et al. 2009).  Combining these data with 
the extensive set of high--angular--resolution HI and CO maps
available for this sample (\S 3.4), enables investigation of the 
spatially-resolved correlation between the SFR surface density and the 
gas surface density (Kennicutt 1998b, Kennicutt et al. 2007), 
variations in the star formation efficiency as functions of
local dynamical environment within disks (e.g., Leroy et al. 2008),
and the physical nature of the apparent star formation
thresholds at low SFR densities  in disks (e.g., Martin \& Kennicutt
2001). 

\subsubsection{The Radio-IR Correlation}

There is a close empirical correlation between the cm-wavelength,
synchrotron-dominated emission and the
FIR luminosity, which also traces (young) massive star
formation (Helou et al. 1985).  As a result the radio continuum 
emission of galaxies is frequently applied as a dust-free tracer
of recent star formation, both for local and distant galaxies (e.g., Condon 1992).
However, it is entirely unclear how presumably 
unrelated physical processes affecting the  
propagation of CR electrons and the heating of dust grains work  
together to yield a nearly ubiquitous correlation between the radio  
and FIR emission over many orders of magnitude.

{\it Spitzer} revealed that the spatially resolved 
correlation between 70\,\micron\ and 
non-thermal radio emission within galaxies shows an `age effect': the cosmic ray 
electron population of galaxies having more intense star
formation largely arise from a recent episode of enhanced star
formation activity and have not had time to diffuse significant 
distances (Murphy et al. 2008).  This work will be extended with {\it Herschel} by 
enabling the study of galaxies over a much larger range of spatial scales and 
physical parameters than possible so far (Murphy et al. 2011).  
These comparisons will be
carried out using 20\,cm maps obtained with the Westerbork Synthesis
Radio Telescope (WSRT; Braun et al. 2007), and with new observations obtained
with the Expanded Very Large Array, The Australia Telescope Compact 
Array (ATCA), the Effelsberg 100\,m Radio Telescope,
and the Robert C. Byrd Green Bank Telescope (GBT).  The goal of these studies is to 
enable the understanding of the underlying physics of radio-IR
correlation on scales down to $\sim$40~pc.

\subsection{The Inventory of Cold Dust in Galaxies}

The {\it ISO} and {\it Spitzer} observatories made major strides toward compiling
a comprehensive inventory of interstellar dust in galaxies.  Nevertheless
our knowledge has been limited by the relatively few spatially resolved 
observations of nearby galaxies at wavelengths above 200\,$\mu$m,
and by the limited sensitivity of those data.
Because of the Planck function and the steep decrease in the opacity of dust
at longer wavelengths (approximately as $\lambda^{-\beta}$, with
$1 < \beta <3$ in most cases), cooler dust can radiate
radiate relatively little power but still account for most of the mass. 
As a result the amounts and distributions of colder dust in galaxies
(T$<$15~K) are poorly measured, and systematic uncertainties of factors of 
two or more in the dust masses of galaxies are not uncommon (e.g.,
Bendo et al. 2003; Galametz et al. 2011).  A combination
of {\it ISO}, {\it Spitzer}, {\it Planck}, and 
ground-based submillimeter measurements of a handful of 
galaxies show evidence for the existence of extended cold dust, with temperatures
as low as 4--6~K (e.g., Hinz et al. 2004, 2006; Meijerink et al. 2005;
Galliano et al. 2003; Dumke et al. 2004; Planck Collaboration 2011a, b).
On the other hand, analysis of a large sample of SINGS galaxies by Draine et al. (2007)
concluded that very cold (T$<$10K) dust contributed no more than half 
of the total dust mass.  The FIR and submillimeter maps of galaxies obtained
with {\it Herschel}, when combined with ground-based 850\,$\mu$m and 
1100\,\micron\ imaging from LABOCA, MAMBO-2, and SCUBA-2
provide an unprecedented opportunity to map and quantify 
the cooler dust emission (e.g., Gordon et al. 2010; Meixner et al.
2010).  Spatially resolving the dust emission
also will allow us to properly weight the contributions of the
different luminosity components (often with very different SEDs)
to the integrated emission, and thus constrain any possible biases
in deriving dust masses and other parameters.

A related problem revealed by {\it ISO} and {\it Spitzer} was the presence 
of strong submillimetre excesses in the SEDs
for some galaxies (e.g., Lisenfeld et al. 2002; Galliano et al. 2003; 
Israel et al. 2010; Bot et al. 2010; Galametz et al. 2011; Planck Collaboration
2011b).  If this excess were associated with a component
of very cold dust this material could comprise the dominant dust component
by mass in the galaxies, but other possible explanations include emission
from (warmer) amorphous dust grains, emission from spinning dust, or
free-free radio emission.  Previous detections of this submillimeter
excess emission in galaxies have been difficult to interpret because 
the beam sizes of the measurements have often been comparable to the
sizes of the entire galaxies.  The sensitivity and
the sub-kpc spatial resolution of the {\it Herschel} SPIRE observations,
when combined with longer wavelength observations mentioned above, should make it
possible to isolate the sources of the submillimeter-excess emission, confirm the
physical nature and origins for the emission, and thereby constrain much more accurately 
the cold dust masses of the galaxies.   The multi-wavelength KINGFISH/SINGS 
maps will also be used to search for extraplanar emission which might be associated 
with starburst driven winds or galactic fountains.

The {\it Herschel} submillimeter continuum maps will also shed light
on the issue of the constancy of the ratio of CO rotational 
line brightness to molecular hydrogen surface density, the notorious
CO/H$_{2}$ X-factor problem (Bollato et al. 2008).  
Comparisons of the dust surface densities of KINGFISH galaxies to
the corresponding column densities of HI and H$_2$ (the latter derived
from CO measurements) provide measurements of the local dust/gas ratio,
over ranges of 2--3 orders of magnitude in gas surface density and
a factor of 20 or more in metallicity (e.g., Draine et al. 2007).
If there are regions in the galaxies where the CO emission severely
under-estimates the total molecular gas column, they may be apparent
as an anomalously high dust/gas ratio (e.g., Leroy et al. 2007, 2009, 2011). 
Our tests will make for an excellent comparison to {\it Fermi} observations which
hope to explain the radial gradient in the Galaxy's diffuse gamma-ray
emission.  The leading candidate is a radial variation (factor of 5 to
10) in the X-factor (Strong et al. 2004).  
Using dust to trace H$_2$ has its own biases and systematics. Because
KINGFISH covers a wide range of environments and this method has
different biases than dynamical, diffuse gamma ray, or absorption-line
approaches it can provide important constraints on the behavior of the
CO-to-H$_2$ conversion factor that complement these other methods.
%Such tests are not likely to provide a definitive calibration for
%variations in the X-factor, but they should be able to place limits
%on gross under-estimation of the molecular gas in galaxies, as 
%suggested by Madden et al. (1997).  

\subsection{The Energy Balance of the Star-Forming ISM}

The spectral coverage of the PACS instrument includes several of the most
important cooling lines in the atomic and ionized ISM, most notably
[CII]157.7\,$\mu$m, [OI]63.2\,$\mu$m, [OIII]88.4\,$\mu$m,
[NII]121.9\,$\mu$m and 205\,$\mu$m.  Even with this limited set of lines
the range of astrophysical applications is broad, 
including mapping of the cooling rates and derived UV radiation intensities
in active star-forming regions and the more quiescent ISM, testing
and calibrating fine-structure lines such as [CII]\,158\,$\mu$m as
a star-formation diagnostic, and constraining the metal abundance scale
in HII regions.

\subsubsection{Cooling of the Interstellar Medium}

The [CII]$158\mu$m and [OI]$63\mu$m far-infrared lines
dominate the cooling of the warm, neutral medium in normal
galaxies.  Photoelectrons liberated from dust grains by UV photons provide the
heat input for the gas (e.g., Hollenbach \& Tielens 1999).  The heating
is relatively inefficient ($0.1-1\%$), and is determined mainly
by the ratio of UV radiation field to density ($G_0/n$).
Observations of the far-infrared cooling lines in representative samples of
local star-forming galaxies were pioneered with {\it ISO} (e.g., Malhotra et al.
1997,2001; Helou et al. 2001; Contursi et al. 2002; Brauher et al. 2008).  
These studies clearly showed
that (1) while the [CII]$158\mu$m emission dominates the line cooling, the
ratio of [CII]/FIR decreases by more than an order of magnitude (from 0.004 to
less than 0.0004) for galaxies with high luminosity and/or warm dust
temperatures, (2) the [OI]$63\mu$m/[CII]$158\mu$m flux ratio ranges from
$0.2-2$, with [OI] taking over as the primary coolant for the high luminosity,
warm sources, and (3) the [CII]/PAH ratios show no such trends with dust
temperature or luminosity.  These results are broadly consistent with a model
that explains the ISM having a diffuse or cirrus-like component, where [CII]
dominates the cooling and the smallest grains provide the bulk of the heating,
together with an ``active" component, where [OI]63\,\micron\ is stronger
than [CII]158\,\micron.

%The mechanism responsible for heating these ``active" regions is uncertain,
%but shock heating is one candidate (Flower \& Pineau Des For\^ets 2010).  
%Unfortunately the poor spatial
%resolution of {\it ISO} and the small samples of detected sources make 
%direct comparisons to models extremely difficult 
%(e.g., Kaufman et al. 1999; Contursi et al. 2002).  With the addition of
%{\it Herschel} observations, the cold dust, PAHs, and the emission-line 
%coolants will be mapped at comparable spatial resolutions, and 
%trace for the first time the detailed heating and cooling in both active and 
%quiescent regions. This will allow us to critically address the 
%importance of small grains in heating the quiescent ISM.
%
%Near bright star forming regions stellar photons from massive (OB)
%stars are likely to dominate the gas heating through the photoejection
%of electrons from grains. However, shocks, dissipation of turbulence,
%and cloud collisions may also contribute in isolated regions. The {\it Herschel}
%observations, together with photodissociation region and shock models (Kaufman et al. 2006),
%map out the dominant heating processes in nuclear, arm, and
%interarm regions. The KINGFISH sample purposefully includes low-luminosity AGNs (about 27\%),
%mostly classified as {\small LINER}s.  These will 
%make it possible to evaluate the impacts of non-stellar activity
%on the dust grain population and on the cooling of the circumnuclear
%gas, dissipating extra energy injected by X-rays and cosmic rays,
%or by extensive shocks.

Near bright star-forming regions, stellar photons from massive (OB)
stars are likely to dominate the gas heating.  However, shocks, turbulence and
cloud-cloud collisions may also contribute in more isolated
regions (Flower \& Pineau Des For\^ets 2010).  Unfortunately, the poor
spatial resolution of {\it ISO} and the small numbers of detected
sources made comparisons to models 
extremely difficult (e.g., Kaufman et al. 1999; Contursi et al.
2002).  With the addition of {\it Herschel} KINGFISH observations,
the cold dust, PAHs, and the far-infrared cooling lines
will be mapped at comparable spatial resolutions, allowing us to
trace the heating and cooling of the gas and dust in both active
and quiescent regions for the first time in a large sample
of nearby galaxies.  The KINGFISH sample purposefully includes 
galaxies (and subregions of galaxies) with a wide range of star-formation 
activity, as well as a fraction of galaxies hosting low-luminosity 
active galactic nuclei (AGNs) AGN (27%, mostly classified as
LINERs from their optical nuclear spectra).  This will
make it possible to evaluate the impact of non-stellar
activity on the dust grain populations (e.g., Smith et al. 2007)
and on the cooling of the circumnuclear gas in the
presence of X-rays, cosmic rays, and/or large-scale shocks
(Meijerink et al. 2007).

\subsubsection{The Range of Physical Conditions of the ISM} 

{\it Herschel} spectroscopic observations are well suited for revealing 
the physical conditions and energetics of the neutral
ISM in these galaxies.  The PACS line observations 
map the emission of the main [CII] and [OI] cooling lines, as well as the 
weaker [NII]122\,\micron, [OIII]88\,\micron, and (in strong sources) 
[NII]205\,\micron\ lines, with linear resolutions of $\le$300\,pc. 
This is comparable to the sizes of typical
star-forming complexes, and the scales over which the formation of
molecular clouds are thought to be triggered.  The [NII] and [OIII] lines probe the 
ionized gas in the obscured, star forming complexes and allow for an estimate of
the gas density and the effective temperature of the ionizing stars.  
The [NII]122/[NII]205 line flux ratio is a sensitive probe of gas density,
for densities of $10 < n_{e} < 1000$ cm$^{-3}$, lower than that probed with the
mid-infrared [SIII] lines seen with {\it Spitzer}.  The combination of spectral line
information from {\it Herschel} and SINGS mid-infrared spectra in the 10--40\,$\mu$m region
with {\it Spitzer} will provide critical measurements of the physical conditions
in the ISM--- temperatures, densities and pressures, 
local UV radiation strength and hardness, and constraints on the clumping
of the gas on scales much smaller than the beam resolutions.

As with nearly all of the investigations being carried out in
KINGFISH most of the diagnostic power of the spectroscopy is exploited
in combination with the matching {\it Spitzer} observations, in this case
with the mid-infrared spectroscopy.  As an illustration of this synergy, 
Figure 3 plots each of the detectable
transitions in terms of ionization potential and critical density.

\subsubsection{Tracing Star Formation with Cooling Lines}

%The bright far-infrared atomic cooling lines are strong spectral features
%in the redshifted submillimeter spectra of distant star-forming galaxies,
%and new facilities such as the Atacama Large Millimeter Array (ALMA)
%are expected to detect these lines in
%large numbers of high-redshift dusty galaxies (e.g., Walter et al. 2009).  
%This capability has focussed
%strong interest on using these lines, in particular the [CII]\,158\,$\mu$m 
%cooling line, as a quantitative SFR tracer.  This question was addressed
%to a limited extent by {\it ISO}.  

Understanding how to use the [CII]158\micron\ (and other FIR fine structure
lines) to estimate the SFR in galaxies is timely and important.
These features are already being observed in some luminous 
star-forming galaxies at high-redshift, with a variety of 
instruments on the ground and in space.
The ability to detect the [CII]  line (as well as [OI] and [OIII]) out
to extremely high redshifts and measure the SFR in extremely young
galaxies is a key science driver for the Atacama Large Millimeter Array (ALMA).
This capability has recently been 
highlighted by the kpc-scale image of the redshifted [CII]  line in the $z = 6$
QSO SDSS J114816.6+52 obtained with the IRAM Plateau de Bure millimeter
interferometer (Walter
et al. 2009). This object has an implied SFR of nearly 2000\,M$_\odot$\,yr$^{-1}$.
Similarly, detection of the [CII] line in
the z = 1.3 ULIRG MIPS J142824.0+352619 (Hailey-Dunsheath et al. 2010)
and in a dozen more galaxies at
$1 < z < 2$ (Stacey et al. 2010) have provided evidence for massive,
galaxy-wide starbursts at early epochs.

Before these cooling lines are applied indiscriminantly as 
SFR tracers, it is important to test their accuracy and reliability
using nearby galaxies where the SFR can be accurately measured
using other methods.  Observations of M31 by
Rodriguez-Fernandez, Brouillet, and Combes (2006) show that the
[CII] emission follows both the H$\alpha$ and {\it Spitzer} MIPS 24$\mu$m 
emission, suggesting its use as a star-formation diagnostic, at least on small
scales.  A calibration of the [CII] line as a global SFR tracer was
derived by Boselli et al. (2002).  Their calibration was limited to normal
late-type galaxies, but even then the derived SFRs had uncertainties
of roughly a factor of ten.  Comparisons
of the integrated [CII] and FIR luminosities of a more diverse sample
of galaxies (e.g., Malhotra et al. 1997; Gracia-Carpio et al. 2011) show 
that this correlation
sometimes breaks down even more severely, particularly
in the regime  where [CII] no longer dominates the cooling,
or when starlight from evolved populations dominates the grain
photoelectric heating.
By mapping the key FIR cooling lines
in the KINGFISH sample we will not only shed light on the range of
physical conditions within normal galaxies,
but we will also establish a critical local baseline against which to
compare the observations of high-redshift
galaxies in the coming decade.

\subsubsection{Metal Abundances and the Nebular Abundance Scale}

The KINGFISH spectroscopy includes measurements of the [OIII]\,88\micron\ line,
a powerful cooling line for HII regions around early-type O stars.
Combining the KINGFISH maps in this line 
with matched-aperture optical spectra from SINGS and the literature will
help resolve the longstanding discrepancy between the 
HII region metal abundance scale as calibrated by conventional 
optical spectroscopy based on auroral-line electron temperatures,
{\it vs} those derived from HII region models fitted to strong lines, 
{\it vs} those measured from optical recombination lines
(e.g., emission lines of OII that are produced by recombination from
OIII).  Discrepancies of up to factors of
2--3 between these calibrations remain present today  (e.g., 
Kennicutt et al. 2003; Perez-Montero \& Diaz 2005; Kewley \& Ellison 2008).
Various explanations for the discrepancies have been offered, 
including the possible presence of temperature fluctuations which
would bias the auroral line measurements; however recent
comparisons of nebular and stellar temperature scales appear
to point more in the direction of systematic errors in the
model-derived abundance scales (Bresolin et al. 2009).
The [OIII]\,88\micron\ line, which
arises from the ground level of the optical [OIII]\,4959,5007\ transitions,
provides an independent measurement of the O$^{++}$ abundance that
is much less sensitive to temperature fluctuations
(e.g., Rudolph et al. 2006).  Calibration of
the density dependence of the emission is provided by the SINGS spectroscopy
of the [SIII]\,18.7\micron\ and [SIII]\,33.5\micron\ lines, an excellent
example of the power of combining {\it Spitzer} and {\it Herschel} spectroscopy.

\section{OBSERVATIONS}  

\subsection{The KINGFISH Galaxy Sample}

The scientific potential of KINGFISH derives
from combining the power of deep infrared imaging with spectroscopy
of the key diagnostic lines, and the vast ancillary data heritage of SINGS. 
To maximize the benefits the KINGFISH imaging and spectroscopy 
have been closely tailored to the existing SINGS observations.
 
The SINGS sample comprises 75 galaxies
within 30 Mpc, selected to cover the range of galaxy types,
the range of galaxy luminosities and masses within each type, 
and the range of dust opacities (as traced by the ratio
of IR to visible luminosities) within this type-mass space
with a strong bias toward star-forming galaxies (see Kennicutt 2003 for details).
%Within the limits imposed by the finite sample size the galaxies
%were also selected to provide extensive parametric coverage in 
%metallicity, gas mass fraction, atomic vs molecular gas fraction, SFR, 
%and bulge/disk/bar structure.  The sample was chosen to reflect the 
%the diversity of the local galaxy population, rather
%than being complete or representative; however the sample does
%include most of the members of the M81 group, and thus forms
%a small population-representative subsample.

The KINGFISH survey did not need to include all 75 galaxies in 
the SINGS sample.  Ten galaxies had already been allocated
extensive observing time as parts of {\it Herschel} Guaranteed Time
programs.  Careful examination of the multi-wavelength data available
the 65 remaining galaxies allowed us to reduce the KINGFISH sample by a 
a further eight galaxies.  The galaxies excluded (NGC 24, 1566, 4450, 4452,
5033, IC 4710, Ho IX, M81dwA) have physical properties and infrared
emission properties which were very similar to other galaxies in
the sample.  At the same time we added four other nearby galaxies which
were drawn from {\it Spitzer} surveys other than SINGS, and which augmented
significantly to the range of physical properties 
covered by the KINGFISH sample.
M101 (NGC\,5457) is one of the largest spirals in the local
volume and exhibits one of the largest  metal abundance gradients 
with galactocentric distance found in a nearby disk galaxy.  
IC\,342 is one of the nearest
giant spiral galaxies and exhibits an unusually large contrast
in star formation properties between a dense starburst nucleus
and a very extended disk.  NGC\,3077 is a peculiar, starburst  galaxy
in the M81 group, that helps complete our coverage of that group.
Finally NGC\,2146 is an unusually dusty early-type galaxy, one
of the nearest LIRGs, and the most IR-luminous galaxy in the sample.

The properties of the KINGFISH galaxies are summarized in 
Table 1.  The values tabulated are much improved thanks to 
new observations from SINGS and elsewhere, and supercede those 
published earlier by Kennicutt et al. (2003).  Recession velocities,
morphology, and projected sizes
are from NED, the NASA Extragalactic Database\footnote{The NASA/IPAC
Extragalactic Database (NED) is operated by the Jet Propulsion
Laboratory, California Institute of Technology, under contract
with the National Aeronautics and Space Administration.}.
Sources for the the other data are given in the extensive 
set of table notes, and are summarized below.

Column (1):  Galaxy identification

Column (2):  Heliocentric radial velocity, from NED

Column (3):  Morphological type, from NED

Column (4):  Approximate major and minor diameters in arcminutes, as listed
   in NED.  These are approximately consistent with RC3 D$_{25}$
   diameters (de Vaucouleurs et al. 1991).

Column (5):  Adopted distance in megaparsecs.  

Column (6):  Method used for adopted distance.  
The distance for each galaxy was based on a
redshift-independent indicator, using, in order of preferences:
Cepheid variable stars (Ceph), tip of the red giant branch stars (TRGB), surface
brightness fluctuations (SBF), Type 2 plateau supernovae (SNII), 
Tully-Fisher relation (TF), and brightest stars (BS).  For
M81DwB we used the mean distance of the M81 Group (M81G --- 3.6 Mpc), as no other
distance measurement is available for this galaxy.

Column (7):  Reference for distance measurements, as documented
in the table notes.
The source of the references is the compilation contained in NED,
and we have attempted to draw the distances from
the largest compilations we could find, in order to minimize the number of
different sources and reduce inconsistencies among different distance
measurements. 

Column (8):  Nuclear type, when available, as derived from optical emission line 
diagnostic plots constructed from nuclear spectra.  These classifications 
come from Moustakas et al. (2010) or (denoted by asterisks)
 Ho, Filippenko \& Sargent (1997). 
The abbreviation SF denotes an HII region-like spectrum, and AGN an 
accretion disk spectrum (LINER or Seyfert).  Readers should note that
although a considerable number of KINGFISH galaxies show non-stellar
nuclear emission most of these are low-luminosity AGNs.  With the
exception of NGC\,1316 = Fornax A none of the galaxies hosts a 
dominant AGN.  Galaxies without a listed
nuclear type either lack nuclear emission lines, have completely obscured
nuclei in the visible (e.g., NGC\,1377), or in the case of most of the
dwarf galaxies lack discernable nuclei altogether.  

Columns (9) and (10):  Mean disk logarithmic oxygen abundance.   
We have adopted the disk-averaged oxygen abundances ($12 + \log O/H$)
from Table 9 of Moustakas et al. (2010).  These authors list
two values for each galaxy: one from the theoretical calibration
of Kobulnicky \& Kewley (2004; denoted KK) and the other from the empirical
calibration of Pilyugin \& Thuan (2005; denoted PT).  Both sets
of measurements are listed in the table.  
We refer the reader to Moustakas et al. (2010) and the papers
cited in \S2.3.4 for a full discussion of the bases and uncertainties
inherent in the two abundance calibrations.  For galaxies lacking
HII region measurements the oxygen abundances are estimated from the
luminosity-metallicity relation (as denoted by $+$ symbols).
For the four galaxies that were not originally in the SINGS sample (IC\,342, 
NGC\,2146, NGC\,3077, and NGC\,5457), we have used values 
from the literature:   (B) Bresolin et al. (2004); (C) Calzetti et al. (2004);
(E) Engelbracht et al. (2008); (P) Pilyugin et al. (2004).

Column (11):  Luminosity at 3.6~$\mu$m 
(a proxy for stellar mass), as derived from flux densities published in 
Dale et al. (2007, 2009) and Engelbracht et al. (2008) and the distances 
listed in Column (5).  

Column (12):  Total infrared luminosity in the range 3--1100~$\mu$m,
L$_{TIR}$,  are derived from the flux densities at 24, 70, and 
160~$\mu$m listed in Dale et al. (2007, 2009) and Engelbracht et al. (2008)
and the distances listed in Column (5).  Equation (4) of Dale \& Helou (2002)
was used to derive L$_{TIR}$ from the fluxes at 24, 70, and 160~$\mu$m.

Column (13):  Star formation rate, in units of M$_{\odot}$~yr$^{-1}$, 
derived using the combination of H$\alpha$ and 24~$\mu$m emission 
as calibrated in Kennicutt et al. (2009), and tabulated in Calzetti et al. (2010),
adjusted to the distances Column (5).  For NGC\,5457, the H$\alpha$
luminosity is from Kennicutt et al. (2008) and the 24\,$\mu$m luminosity from
Dale et al. (2009).

Column (14):  Total stellar mass in units of solar masses, derived
using the method of Zibetti et al. (2009), and published by 
Skibba et al. (2011), but rescaled when necessary to the distances
adopted here.  

Column (15):  Classification of the galaxy as a high surface brightness
(B) or faint surface brightness (F) system for the PACS and SPIRE
observations, as described in \S 3.2.

Figure 4 illustrates the distribution of galaxy types and SFRs 
of the galaxies in the KINGFISH sample, plotted in both cases as
functions of distance, to give a sense of the linear resolution
available for specific classes of galaxies.  As expected, the distribution 
of morphological types is fairly independent of distance, except for the 
irregular galaxies, which are overwhelmingly clustered in the lowest 
distance bin on account of being mostly faint galaxies (Table~1). There is 
also a weak trend for earlier type galaxies to populate larger distance bins, on 
account of these types usually being larger, more luminous galaxies. 
Similarly, low SFRs are generally found in the lowest distance bins.  
Most of the closest low--SFR 
galaxies are gas--rich dwarfs.  Unusual types of galaxies, for example
early-type galaxies with high SFRs, tend to lie at larger distances, 
because they are rare objects and only appear when one probes a larger
cosmic volume.  

In terms of other properties the KINGFISH sample covers nearly two orders of 
magnitude in oxygen abundance ($7.3 < 12 + \log(O/H) < 9.3$, 
four orders of magnitude in 3.6~$\mu$m 
luminosity ($\sim$10$^6$--10$^{10}$~L$_{\odot}$), and in 
total infrared luminosity ($\sim$4$\times$10$^6$--10$^{11}$~L$_{\odot}$), and 
a similar total range in SFRs ($\sim$0.001--7~M$_{\odot}$~yr$^{-1}$). 
 
Figure~5 shows the distribution of the KINGFISH galaxies in terms
of the well known bimodal relation between specific SFR (SSFR) and
stellar mass.  Most of the galaxies populate the star-forming 
``blue cloud", which reflects the deliberate survey bias to actively
star-forming systems.  However higher-mass galaxies with low SFRs
consistent with the red sequence are also represented in the sample,
though they tend to be early-type spirals rather than E--S0 systems.
The sample lacks examples of the most massive red galaxies.  
Also apparent is the lack of clear separation in SSFRs between
early-type and late-type galaxies.  This again reflects the design
of the SINGS and KINGFISH samples; we deliberately included galaxies
with a wide range of infrared luminosities independent of morphological
type, so the sample contains unusually large numbers (relative to
any volume-limited sample) of peculiar early-type galaxies, often
with relatively high infrared luminosities and SFRs.

\subsection{PACS and SPIRE Imaging}

%\newcommand{\mjysr}{MJy\,sr$^{-1}$}
%\newcommand{\muuu}{$\Sigma_{160 \mu {\rm m}}$}
%\newcommand{\ropt}{R$_{25}$} 
%\bibitem[Boselli et al.(2010)]{2010PASP..122..261B} Boselli, A., et al.\ 2010, \pasp, 122, 261 
%\bibitem[Nguyen et al.(2010)]{2010A&A...518L...5N} Nguyen, H.~T., et al.\ 2010, \aap, 518, L5 

A subsample of 55 galaxies are being imaged 
with PACS$+$SPIRE to well beyond the optical radius at
70, 100, 160, 250, 350, and 500\,\micron; the remaining 6 KINGFISH galaxies, 
which are also part of the {\it Herschel} Reference Survey Guaranteed Time Key Project 
(Boselli et al. 2010a), are imaged with PACS alone. 
Our map sizes are designed to probe cool dust, if present, outside the optical 
disk and to provide sufficient sky coverage to remove instrumental artefacts; 
we thus acquire for all galaxies with both instruments square maps of 
dimension at least 1.5 times the optical diameter.
A minimum map size of 10$\times$10\,arcmin is imposed in order to maximize
efficiency, given the observing overheads, and better standardize observations.
At the SPIRE wavelengths, sensitivity is limited by confusion even for modest
exposure times. On the other hand, at the shortest PACS wavelengths we cannot
easily reach the confusion limit, so we chose instead to obtain
sufficient depth to detect the diffuse disk emission within 80-90\% of \ropt,
and to detect individual IR sources with high signal/noise beyond \ropt.
The goal is to map the diffuse emission at PACS fainter surface brightness
levels through a strategy of smoothing and combination with the SINGS 70 
and 160\,\micron\ MIPS images, which are much deeper on large angular scales.

The {\it Spitzer} MIPS images for these galaxies show that the sample spans a wide
range in FIR surface brightnesses at the optical radius \ropt\ 
(Munoz-Mateos et al. 2009).
We therefore evaluated the 160 \micron\ surface brightness \muuu\ 
at \ropt, and analyzed the histogram of the resulting \muuu\ values.
This enabled us to divide the sample into two
groups, a ``bright'' (high FIR surface brightness at \ropt) group with median 
\muuu$\sim3$\,MJy\,sr$^{-1}$ and a ``faint'' one (low surface brightness) 
with \muuu$\lesssim 1$\,MJy\,sr$^{-1}$.  These two subsets are listed
as B and F, respectively, in Column (15) of Table 1.
Our observations were devised to achieve these
1$\sigma$ per-pixel sensitivities at \ropt\ at 160\,\micron.
By adjusting the exposure times between the two groups we were able
to optimize the requested time allocation.

The SEDs of most of the KINGFISH galaxies peak in the 75$-$170\,\micron\
region and fall off rapidly at longer wavelengths (Figure 2). Since the
primary goal of our imaging is to accurately map the distributions of
dust mass and temperature, we need to tailor our exposure times to provide 
comparable signal/noise for a typical galaxy SED 
across this wide range in wavelength.  For most
of the galaxies, this required an estimated extrapolation of the brightness
distributions into the submillimeter. This was done using the MIPS \muuu\
at \ropt\, together with empirical or model SED estimates (Dale et al. 2007).

To maximize efficiency, we have adjusted our exposure times to achieve
comparable sensitivity across the range in wavelengths spanned by
PACS$+$SPIRE, and better constrain the distributions of dust mass
and temperature in the outer regions of the galaxies. The PACS maps
are acquired in scan mode at medium speed of 20\arcsec\,s$^{-1}$, with homogeneous 
coverage and the array oriented at a 45$^\circ$ angle relative to the scan
direction.  Each PACS blue wavelength observation (70\,$\mu$m and 100\,$\mu$m)
consists of two chained AORs, with perpendicular scans to minimize detector
anomalies.  Four AORs per galaxy are required for complete PACS coverage, 
since PACS acquires a single blue wavelength simultaneously  
with the red one (160\,\micron).  Hence an observation consisted of 
two observations at 70\,$\mu$m and 100\,$\mu$m and four at 160\,\micron.
These prodecures were also adopted to filter out transient phenomena 
(e.g., asteroids).  Each AOR consisted of three scan repetitions for
bright targets and six repetitions for faint targets.

The SPIRE observations were performed similarly, using the Large Map mode
at the nominal scan speed of 30\arcsec\,s$^{-1}$, with the array oriented 
in two orthogonal scans
at 42.4$^\circ$ and -42.4$^\circ$ relative to the scan direction.
Galaxies were observed in a single AOR with two and four repetitions for
the bright and faint group, respectively.  Table 2 lists pixel sizes
and approximate sensitivity limits for each wavelength in the PACS
and SPIRE observations.  Note that at the longest wavelengths the observations
are confusion limited (Nguyen et al. 2010).

%The observations for faint sources 
%achieve 1$\sigma$ per-pixel sensitivities of
%5.0, 5.0, 2.2, 1.1, 0.6, 0.4 \mjysr\ at 
%70, 100, 160, 250, 350, and 500\,\micron, respectively
%(equivalent 1$\sigma$ point-source sensitivities are 
%3.6, 4.1, 5.6, 4.5, 3.8, 5.4\,mJy, respectively).
%xxx These are taken from HSpot, not from our images; need to update these 
%numbers. xxx
%The {\it bright} source 1$\sigma$ sensitivities are a factor of
%$\sqrt{2}$ brighter. 
%SPIRE sensitivities are slightly below the confusion limit
%(Nguyen et al. 2010).

For IC\,342, the largest galaxy in our sample, we used the Parallel Mode,
in which PACS and SPIRE observations are acquired simultaneously.
Again, we required complete PACS coverage, so we used two repetitions at slow speed
(20\arcsec\,s$^{-1}$) in
orthogonal directions for each PACS blue wavelength, in order to better remove
striping, 1/f noise, and mitigate beam smearing along the scan direction.
Estimated sensitivities are given in Table 2.

%These observations achieve 1$\sigma$ sensitivities of
%11.7, 12.3, 5.3, 0.3, 0.2, 0.1 \mjysr\ at 
%70, 100, 160, 250, 350, and 500\,\micron, respectively.
%As for the above observations, SPIRE sensitivities are below confusion.
% These are from HSpot, 21/3/2011.

\subsection{Spectroscopic Observing Strategy}

The KINGFISH spectroscopic observations consist of spectral mapping in
the key diagnostic emission lines [CII]158$\mu$m, [OI]63$\mu$m,
[OIII]88$\mu$m, [NII]122$\mu$m, and (in areas of high surface
brightness) [NII]205$\mu$m.  The observing strategy closely follows that
of the SINGS project, covering nuclei, selected extra-nuclear regions,
and full radial strips out to a limiting surface brightness threshold.
All of the targeted regions have been mapped previously with the {\it Spitzer}
IRS spectrometer.  The galactic centers and extra-nuclear regions were
observed at low resolution with {\it Spitzer} from 5--14 $\mu$m, and with the
{\it Spitzer} high-resolution spectrometer at 10--37 $\mu$m. The radial strips
coincide with low-resolution SINGS radial strip maps covering
14--38$\mu$m.

Utilizing the 47\arcsec$\times$47\arcsec\ PACS spectrometer field of
view, a total of 51 nuclear pointings, 28 strips (covering 115 PACS
fields of view), and 41 extra-nuclear positions in 14 separate galaxies
were targeted.  The selection criteria for the extranuclear regions was
a physically-based strategy to ensure the widest possible coverage in
metal abundance, infrared surface brightness, and 8/24$\mu$m flux ratios
(high 24/8 values are indicative of dust heated by the intense radiation
fields associated with star-forming regions; e.g., Draine \& Li 2007).  
A small number of SINGS targets were dropped entirely
from the KINGFISH spectroscopic sample due to non-detection in {\it Spitzer} MIPS
wavebands.  An example of the PACS + {\it Spitzer} spectroscopic targeting is
given in Figure 6.

Line surface brightness projections were based on nominal scalings from
the total infrared (TIR) surface brightness derived from the SINGS MIPS
photometry, together with average fractional line luminosity
measurements of nearby galaxies from the {\it ISO} Key Project (Malhotra et al.
2001).  Observations of strips were truncated when the peak projected
TIR surface brightness within a single pointing of PACS fell below
10$^{-7}$ W m$^{-2}$ sr$^{-1}$, with additional slight trimming to
adjust for time constraints and requirements on the cadence of
background observations.  Targeted regions (strips, single nuclear
pointings, and extra-nuclear regions) were divided into three bins of
total infrared surface brightness, and line repetitions were chosen
to deliver signal-to-noise of at least 5 for the
median brightness within each bin.  A minimum of one repetition per
raster position was allocated for [CII] in the brightest bin, with a
maximum of 4 repetitions for [OI] and [NII] in the faintest bin.  Where
possible, multiple line observations of a given target were grouped into
a single AOR.  The sensitivity of these maps varies with wavelength
(due to changes in sensitivity across the spectral range), but for
areas with full depth coverage a typical 1-$\sigma$ sensitivity is
$2 - 4 \times 10^{-9}$ W\,m$^{-1}$\,sr.  This noise is measured from
single-pixel extractions (2\farcs85 pixels).  Lower surface brightness
sensitivities can be achieved by averaging over larger regions 
before fitting the spectra.

Since a substantial majority of the sample is comprised of galaxies with
photometric radius larger than the maximum PACS chopper throw angle,
\emph{unchopped} line scan mode was used.  In early tests on two
smaller KINGFISH galaxies for which additional chopping observations
were obtained, line fluxes and sensitivity per unit observing time
compared favorably (within the calibration uncertainty limits) between
maps created with and without chopping.  In individual targeted regions,
a 2$\times$2 sub-pixel dither pattern of 4\farcs5 $\times$ 4\farcs5, 
or 0.48$\times$0.48 PACS spectrometer pixels, was utilized to mitigate
the undersampling of the beam delivered by the telescope, which is
particularly significant at the shortest wavelengths.

For the full radial strips, a dither pattern of
23\farcs5 $\times$ 4\farcs5 was used, both to help recover the
undersampled beam and minimize coverage gaps at the perimeter of the
maps.  This effectively increases the observing time per pixel in most
areas by a factor of four over that of a single raster position.  Since
the radial strip position angles were dictated by preexisting
{\it Spitzer} IRS observations, no attempt was made to adapt the roll angle of
the square PACS spectrometer field to this angle.  In practice this
leads to little or no loss of joint coverage between the {\it Spitzer}
IRS long-low scan maps and the corresponding PACS maps.

Offset backgrounds fields were specified nearby in areas of low infrared
surface brightness, and were visited at least every 2 hours, which, in
the case of galaxies with long radial strips, required multiple
background visits during individual line scans.

\subsection{Ancillary Multi-Wavelength Observations}

Achieving the main scientific aims of this project requires a strong set
of multiwavelength data extending from the UV to the radio to complement the {\it 
Herschel}
observations.  These provide essential information on the
gas and dust emission at shorter wavelengths than probed by {\it Herschel}, the
stellar populations which heat the dust and ionize the gas, and the
associated cold and warm gas components.  

Most of these ancillary datasets were assembled for the SINGS
project (Kennicutt et al. 2003), or were carried out by other
groups in parallel with the SINGS project.  Briefly the imaging 
data include ultraviolet images (153\,nm and 230\,nm) 
from {\it GALEX} (Gil de Paz et al. 2007), ground--based B,V,R,I, and
H$\alpha$ images, Fabry-Perot \halpha\ maps and velocity fields
(Daigle et al. 2006, Dicaire et al. 2008), 
near-IR J, H, and K imaging compiled from
the 2MASS survey (Jarrett et al. 2003), and of course the {\it Spitzer}
imaging at 3.6, 4.5, 5.8, and 8\,\micron\ (IRAC) and at 
24, 70, and 160\,\micron\ (MIPS).  Radio continuum maps are
available from the WSRT (Braun et al. 2007) and the VLA,
and Chandra X-ray imaging for approximately 75\%\ of the sample
is being obtained or compiled (Jenkins et al. 2011).

Deeper visible and near-infrared broadband imaging of the
northern KINGFISH sample is being obtained in the $g$, $I$,
and $H$, using the wide-field imagers at the Calar Alto Observatory.
These are explicitly designed to be able to produce high-quality
stellar mass maps of the galaxies based on state of the art
stellar populations models (Zibetti et al. 2009).

A number of the galaxies in the KINGFISH sample are near
enough to have deep HST imaging of the resolved stellar populations,
for example through the ACS Nearby Galaxy Survey Treasury
(ANGST; Dalcanton et al. 2009).   Those observations can be used 
to derive spatially-resolved star formation histories 
(e.g., McQuinn et al. 2010a, b; Weisz et al. 2011), and 
strong independent constraints on SFRs and star formation
timescales derived from integrated light.

Spectroscopy includes ground-based optical spectra (3,700--7,000\AA;
Moustakas et al. 2010), 
{\it Spitzer} IRS (low and high resolution, 5--40\,\micron)
and MIPS SED (50--100\,\micron) spectral maps.  The SINGS IRS spectral
maps (low-resolution spectral strips, circumnuclear maps, extranuclear
region maps) are an especially unique resource.  
{\it Spitzer} spectra are obtained with fixed
slits, and one needs multiple-pointing maps to properly 
interpret the combined spectra, especially in nearby galaxies where
sources are extended and often multiple sources are blended in the
beam at the longest wavelengths.  

A number of radio line surveys have targeted subsets of the
SINGS and KINGFISH samples, to study the relationships between
the cold gas, dust, and star formation in the galaxies.  
THINGS (Walter et al. 2008) produced atomic gas (HI) maps of 24
KINGFISH galaxies with $\sim 6\arcsec$ resolution. For another 21
galaxies, we have somewhat lower quality ($\sim 15\arcsec$ resolution)
H\,I imaging from new or archival VLA and WSRT observations. Thus a
total of 45 KINGFISH targets have in-hand atomic gas maps with the
exceptions being southern and early-type galaxies. The HERACLES survey
(Leroy et al. 2009) used the IRAM 30-m telescope to obtain deep,
wide-field CO $J=2-1$ maps of 41 KINGFISH targets.
Other CO $J=1-0$ surveys are being carried out with the CARMA array and
the Nobeyama Radio Observatory, and CO $J=3-2$ observations of
the centers of many of the galaxies are being obtained at the JCMT
as part of the JCMT Nearby Galaxies Legacy Survey.  Likewise radio continuum
observations of the galaxies or regions within the galaxies
are being carried out on a number of facilities, as discussed
already in \S2.1.3 (also see Murphy et al. 2011).  Although
the teams on these projects often include members of the KINGFISH
collaboration most are independent undertakings, and the
corresponding datasets will be produced and released independently
from the KINGFISH project.

\section{Data Processing, Analysis, and Data Products}

As appropriate for a large Key Programme the {\it Herschel} observations
from KINGFISH will be converted to fully documented data products
for archival use by the community.  In this section we briefly describe
the processing and expected nature of the public data products.

Most of the data processing is performed within the {\it Herschel} Interactive 
Processing Environment (HIPE; Ott 2010), with subsequent customized processing
as described below.  The KINGFISH observations present particular challenges
in the data processing, with significant emission over a wide range of
spatial scales, and (for spectroscopy) with target sizes that are large
compared to the standard chopping throw of the PACS spectrometer.  
At the time of submission of this paper the higher-level processing
was still under development.  Here we describe the basic principles
and objectives of the data reduction efforts; more detailed documentation
will be provided as part of the future KINGFISH data releases.

\subsection{PACS Scan Maps}
 
At the wavelengths covered by PACS the KINGFISH galaxies are
expected to emit fluxes on a range of spatial scales, ranging
from unresolved point sources to diffuse emission on scales 
extending to at least an order of magnitude larger than the 
instrumental PSF.  The presence of this extended structure
in the presence of observations with a high thermal background
poses a challenge for the processing of the PACS observations.
Early tests by our group and others reveal that the recovery
of diffuse emission is strongly dependent on the processing
algorithms used.  As a result we are applying more than one
approach to the PACS data processing.  We first describe 
a standard reduction using a slightly modified version
of the HIPE algorithms, and then describe a separate reduction
which combines low-level HIPE processing with a separate
mapping stage using the Scanamorphos package (Roussel 2011).

\subsubsection{Initial HIPE Processing}

The PACS processing mostly follows the recommended standard procedure (i.e. pointing
association, conversion from engineering units to physical units, flat-fielding).
This provides pixel timelines that are almost ready to be reprojected in the sky to
produce a map. 
Prior to this operation, however, one needs to correct the recorded signals
for cosmic ray events and 1/f noise (the latter corrected in the post-Level 1
processing described here).  Anomalies (glitches) are removed using the
so-called second-level deglitching method.
Instead of inspecting bolometer timelines for outliers, each individual
readout is compared to the estimated sky value at the same sky position to
identify outlier values.  This reference is composed of readouts  
from different pixels in the detector, thus it is necessary to first remove the
relative pixel-to-pixel offsets.  This is achieved either by subtracting 
a median image, or by applying a high-pass filter with a very large filtering window.
This method has proven to be more robust than the default pipeline method, in
particular with respect to avoiding rejection of compact sources which can
mimic instrumental artifacts in their timeline signature.
Another advantage of the second-level deglitching is that is also flags pixels that
are severely affected by cross-talk effects, as those are obviously outliers with
respect to the sky value at their pointing position.

Once the data have been deglitched, flat-fielded, and calibrated into physical units,
maps are generated using two different approaches as described below.

\subsubsection{PhotProject maps}

Low-frequency noise is the dominant source of uncertainty in the PACS 
images of extended sources.  Unlike 
SPIRE, the PACS photometer does not sample the temperature of its focal plane with
sufficient accuracy to enable the removal of thermal drifts from the data. Therefore
a strategy using only the signal must be used to separate the
$1/f$ noise from the signal of interest.  This requires the application of
a median high-pass filter to the data before reprojecting the time series scans
back to the sky.  Smaller filtering windows would better sample the
time variation of the thermal drifts, but small filter windows also
risk removing more of the real extended emission from the source.
Therefore this filtering is only applied to the section of the
scans which do not cover the target. We determine the location of the target in
the timeline by projecting a mask, made on a first rough version of the map, into the
timelines using the pointing information. Once this is done, the map is created with
the {\tt PhotProject} task of HIPE that implements the drizzle algorithm. 

Some maps can present a striped pattern appearing in small patches over the
field. This is due to electrical interferences that are picked up by
the readout circuits. The very low-level of these interferences make them very
difficult to dectect and remove, and at this time there is no efficient method
available for eliminating these artifacts entirely. 

\subsubsection{Scanamorphos maps}

Currently the preferred processing of the KINGFISH PACS observations
is carried out using the Scanamorphos
software\footnote{\it http://www2.iap.fr/users/roussel/herschel/}
(Roussel 2011). Its main task is to subtract the brightness drifts caused by the
low-frequency noise (comprising both the thermal drifts of the telescope 
and detectors and the uncorrelated 1/f noise of the individual 
bolometers), before projecting the data onto a changeable spatial grid.
The algorithm employs minimal assumptions about the noise and the signal, and extracts the drifts
from the data themselves, taking advantage of the redundancy built in the scan observations.
With the nominal settings used by the KINGFISH survey, the drifts can be determined on timescales
greater than or equal to 0.7\,s at 70 and 100\,$\mu$m, and 0.9\,s at 160\,$\mu$m (for a sampling
interval of 0.1\,s). These timescales correspond to lengths between 1.5 and 2.5 times the beam FWHM,
from 160\,$\mu$m to 70\,$\mu$m. The second-level deglitching was performed, and the option to detect
and mask brightness discontinuities was also used. The data are weighted by the inverse square
high-frequency noise of each bolometer in each scan.

The output of Scanamorphos is in the form of a FITS data cube for each band. 
The four planes of the data cube are a signal map, error map, a map of 
the drifts that have been subtracted, and the weight map.
Currently there is no propagation of errors associated with the successive 
processing steps in the pipeline.  For each pixel, 
the error is defined as the unbiased statistical estimate of the
error on the mean. The brightness unit is Jy/pixel, and the pixel size is 
one fourth of the beam, as listed in Table 2.

Over the past year our team has invested considerable effort into
quantifying the uncertainties introduced both into integrated photometry
and spatially-resolved mapping of the FIR emission in the PACS maps.
These include intercomparisons of PACS maps produced with different
processing algorithms and independent comparisons with maps produced
by {\it Spitzer} MIPS, {\it IRAS}, and other instruments.  Based on
our preliminary comparisons it appears that the Scanamorphos processing
preserves most of the extended emission in most of the galaxies analyzed
to date, and we plan to deliver these as our primary PACS data products
during at least the early stages of the project.  More documentation
on the processing and uncertainties in the maps will be provided
as parts of those future deliveries.

\subsection{SPIRE Scan Maps}

The raw KINGFISH SPIRE data are processed through the early stages of
HIPE to apply all of the standard corrections to Level 1 (including
deglitching), and to convert the data to physical units of
flux density.  A line is fit to the data for each scan leg after
masking out the galaxy, and this fit is subtracted from the data.
Discrepant data (usually a rogue bolometer, of which there are typically fewer than
one per map) are also masked, and the data are mosaicked using the naive mapper
in HIPE.  The map coordinates are then adjusted so that the positions of
the point sources (measured using StarFinder; Diolaiti et al. 2000)
match the positions in the MIPS 24~\micron\ images (with an average
correction of $\sim$3\arcsec).  Finally, the images are
converted to surface brightness units by dividing by the beam areas
published in the SPIRE Observer's Manual:  426, 771, and 1626 sq
arcsec at 250, 350, and 500~\micron, respectively.
The output of our pipeline is six simple FITS files for each galaxy,
comprised of a calibrated image and uncertainty map for each of the three bands.

\subsection{PACS Line Maps and Data Cubes}

Readout sample ramps of raw PACS spectral data are fitted onboard the
{\it Herschel Space Observatory}.  The fitted ramps are then calibrated and
processed using HIPE.
After basic calibration and the insertion of instrument status
information into the meta-data associated with an observation, data
samples are separated based upon the observed grating positions.  Each
observed line is processed independently from other lines observed
during the same AOR.  Because motions of the telescope can affect the
baseline level of a pixel, individual raster positions are treated
independently.  When data lack chopping offset fields a canonical dark
image is subtracted from each pointing.

The reliability of unchopped mode observations is significantly enhanced
by the large amount of redundancy in the data.  During each repetition
of an unchopped line scan, each spectral pixel at a given spatial
position samples precisely the same wavelength 20 times (in two sets of
up-down grating scans of 75 steps per scan, with 5 readouts per grating
step).  The 16 individual spectral pixels at each spatial position
together sweep out a slightly larger wavelength range, and further
increase the density of wavelength sampling.  Additional repetitions add
to this redundancy.  This large series of repeated scans is referred
to as a ``data cloud".

Pixel-to-pixel baseline response variations are found to be several
times larger than the predicted continuum levels underlying each line.
Therefore, readout sequences are normalized by subtracting a low-order
polynomial fit to the time sequence of data for each pixel, omitting
those readings which sample the line itself.  All data at a given
spatial position are then binned within a wavelength grid that
oversamples the theoretical resolution by a factor of two.  Each
wavelength interval in the final binning has $\sim$100 individual
samples contributing (per repetition).  Within each bin, data that
deviate from the bin mean by more than three standard deviations are
flagged as outliers and rejected.

Carefully positioned off-source fields, which are free of known infrared sources, are
observed with a cadence of at least one every two hours, using one
repetition of the same line scanning sequence.  Data from these
positions are combined to create a generalized background spectrum.  The
background spectrum at each spatial position is fitted with a low order
polynomial to create a noise-free background map, which is then subtracted
from each raster position in an observation.  Background-subtracted data
are produced for each raster position individually.

Final data cubes for each galaxy are obtained by projecting all raster
positions of a given line onto a common set of coordinates based on
{\it Herschel} pointing information recorded in the meta data.  Pointing
uncertainty is of order 3\arcsec.  At the $\sim$10\arcsec\ resolution
delivered, this pointing uncertainty does not significantly degrade
the image quality.

%For each galaxy, integrated line surface brightness maps and velocity
%maps in all targeted positions are produced by fitting the final data
%cube.  Uncertainty maps are also produced, based on the scatter within
%the trimmed, uncollapsed data cloud, with additional estimates obtained
%from the residual deviations in the continuum of each observed line at
%each position.

Spectral line maps are obtained for each galaxy by applying a
multi-step fitting procedure on the data cube. Gaussian fits are
performed on each spectrum of the cube, after removing a third order
baseline excluding the spectral line region determined by the neutral
hydrogen systemic velocity and velocity range information. This
process yields peak surface brightness, integrated intensity,
velocity centroid, and line-width for each pixel of the cube.

In those pixels where the fit fails to meet a number of validity
criteria (for example, the centroid found is outside the allowed HI
velocity range) the fit is replaced by straight integration. Spectra
are integrated on a range of velocities corresponding to the [CII]
emission at that position, or based on the HI information if the [CII]
results were found to be invalid. Uncertainty maps, based on the 
scatter within the trimmed, uncollapsed data cloud,
 and other diagnostic
information maps are simultaneously produced using error propagation.

As a result of the late commissioning of the PACS unchopped line 
scan mode our data processing algorithms are still under development.  
At present we perform the fitting and integration on spectra with and
without the off position removed, and combine the results using
signal-to-noise criteria. We expect that as the understanding of how
to perform the data deglitching improves this step, and the subtraction 
of the third order baseline will become obsolete and the maps will be 
reliably produced based on the off-subtracted data.  
In the future we will also correct the PACS data for long-term 
transient effects, which are expected with the Ge Ga detectors used
in the spectrometer. Transients with time
constants of a few hundred seconds are very noticeable in all 
detectors after a significant
change in the background level (for example immediately following
a calibration sequence, or after a move to a new position during a raster scan). 
%These effects, which are more pronounced in the red detector array, are
%corrected assuming the transient response has an exponential form
%(Fadda \& Jacobson 2011). 
The correction is performed as part of the 
standard PACS spectroscopy pipeline reduction in future versions of 
the HIPE reduction software (from HIPE 7.0), and we are exploring
our own custom algorithms as well.  Transients caused by cosmic ray
hits on individual detector elements are 
not corrected by this technique, but these events are filtered out in the PACS 
pipeline using an outlier-rejection method which is very effective 
because of the huge degree of
redundancy present in a typical observation with the spectrometer.

\subsection{KINGFISH Data Deliveries}

A staged delivery of complete PACS and SPIRE images and spectral line maps 
is planned as part of the KINGFISH project.  The form of the projects
is described above, and deliveries will be accompanied by full documentation.
Since the time scale for completing the {\it Herschel} observations
is uncertain it is not possible to produce a firm delivery schedule
at this time.  However the first set of KINGFISH data, comprising a
complete set of processed SPIRE maps, was delivered to the HSC in
June 2011, and should be available via the HSC User Reduced Data 
pages.\footnote{http://herschel.esac.esa.int/UserReducedData.shtml}

\section{A Taste of KINGFISH:  Image Atlas, Sample Spectra, and SEDs}

This paper is intended as an introduction and reference paper
for the many science papers which we anticipate will follow from
the KINGFISH team and from archival users of the survey data.
As illustrations of the scientific potential for these data, however,
we present examples of early results from the imaging and spectroscopic
observations.

To illustrate the wealth of information provided by the {\it Herschel}
PACS$+$SPIRE maps, individually and in conjunction with {\it Spitzer} images, 
we present in Figure 7 montages of SINGS and KINGFISH images for all 61
galaxies in the KINGFISH sample.  The printed edition shows an example
for NGC\,6946, with the nine panels depicting grayscale images in three
{\it Spitzer} bands (3.6, 8.0,
and 24\,\micron), the three {\it Herschel} PACS bands (70, 100, 160\,\micron), 
and the three SPIRE bands (250, 350, 500\,\micron).  Images for all 61
galaxies can be found in the on-line version of this article.  At the
time this article was accepted PACS imaging had not yet been completed
for NGC\,584.

As a result of the late commissioning of the full PACS unchopped line scan
mode (two years after launch) our spectroscopic observations are still
in progress, and our data processing algorithms are still 
at an early stage of development.  Nevertheless 
the powerful capabilities of these observations are illustrated in 
Figures 8 which show preliminary [CII], [OI], [NII], and [OIII] 
line maps and a velocity map for NGC~3521.  Figure 9 shows spectra
from representative regions that have been extracted from these maps.

We conclude with an illustration of how the
sample size and diversity of the KINGFISH survey and the richness
of its ancillary dataset can help to
clarify and unravel some of the astrophysical questions raised
in the first part of this paper.  

Early studies of {\it IRAS}-selected samples had suggested that local
galaxies tend to possess higher mean dust temperatures than high--redshift
galaxies. This view has been slowly changing over the past decade, as
more and more sub--mm data have been accumulating for significant
local galaxies samples (e.g., Dunne \& Eales 2001; Galliano et al. 2003, 
2005; Dumke et al. 2004), 
thus extending the wavelength range to a regime that can probe
the cold dust population and avoiding the earlier selection biases.
Recent studies using submillimeter selected samples from {\it Blast} (Dye et al.
2009), {\it Herschel} (Boselli et al. 2010) or {\it Planck} (Planck Collaboration
2011) find that nearby galaxies exhibit colder temperatures than what
was previously determined. The sub--mm excess of the local galaxies
also is found to be more pronounced in metal--poor dwarf galaxies (e.g.,
O'Halloran et al. 2010, Galametz et al. 2010, 2011). Although the  
nature of the
sub--mm excess in low metallicity galaxies is not yet fully known
(e.g., Bot et al. 2010),
studies of large samples such as that of Galametz et al. (2011), that
combine infrared and sub--mm data, offer the best opportunity to investigate
the systematics of the sub-mm excess emission.

The early {\it Herschel} KINGFISH data (Dale et al. 2011)
offer an excellent opportunity
to extend these studies of submillimeter dust by comparing with a consistent set of
observations
the FIR--submillimeter SED shapes of the sample, and test whether
these SEDs correlate systematically with the gross physical properties
of the galaxies.  As an illustration 
Figure 10 shows the dependence of the integrated
160\,$\mu$m/500\,$\mu$m flux density ratio as a function of the mean
metal abundance, TIR luminosity, and TIR/FUV flux ratio (all 
plotted in logarithmic form).  Significant correlations
appear in all three comparisons, though the considerable uncertainties
in individual mean metallicities make that a less convincing correlation
than the others.  Similar systematic trends in SED shape have been seen in
early data from the {\it Herschel} Reference Survey (Boselli et al. 2010).
These results show that the submillimeter emission
{\it increases} as one goes from luminous, dusty and metal-rich
galaxies to the more metal-poor dwarf galaxies, even though the
SEDs of the latter often show indications of more warm dust at
shorter wavelengths.  It is tempting to associate the 160/500\,$\mu$m 
spectral slope as a proxy of the cold dust temperature, in which case
the correlations in Figure 10 would suggest the presence of a larger
amount of cold dust in low-metallicity environments.  It is 
also possible however that the trends seen are produced by a 
systematic change in the emissivity properties of the dust.
In either case the comparison suggests suggest a continuous
trend rather than a dichotomy between the massive spiral
and less massive dwarf galaxies.

\acknowledgments

This research has made use of the NASA/IPAC Extragalactic Database 
(NED) which is operated by the Jet Propulsion Laboratory, California 
Institute of Technology, under contract with the National Aeronautics 
and Space Administration.

%{\it Facilities:} \facility{Nickel}, \facility{HST (STIS)}, \facility{CXO (ASIS)}.




\clearpage

%% the file, uncomment the \plotone, \plottwo, and
%% If you need a layout that cannot be achieved with \plotone or
%% \plottwo, you can invoke the graphicx package directly with the
%% \includegraphics command or use \plotfiddle. For more information,
%% please see the tutorial on "Using Electronic Art with AASTeX" in the
%% documentation section at the AASTeX Web site,
%% http://www.journals.uchicago.edu/AAS/AASTeX.
%%
%% The examples below also include sample markup for submission of
%% supplemental electronic materials. As always, be sure to check
%% the instructions to authors for the journal you are submitting to
%% for specific submissions guidelines as they vary from

%% This example uses \plotone to include an EPS file scaled to
%% 80% of its natural size with \epsscale. Its caption
%% has been written to indicate that additional figure parts will be
%% available in the electronic journal.

\begin{figure}
%\epsscale{.90}
\vspace{-6cm}
\plotone{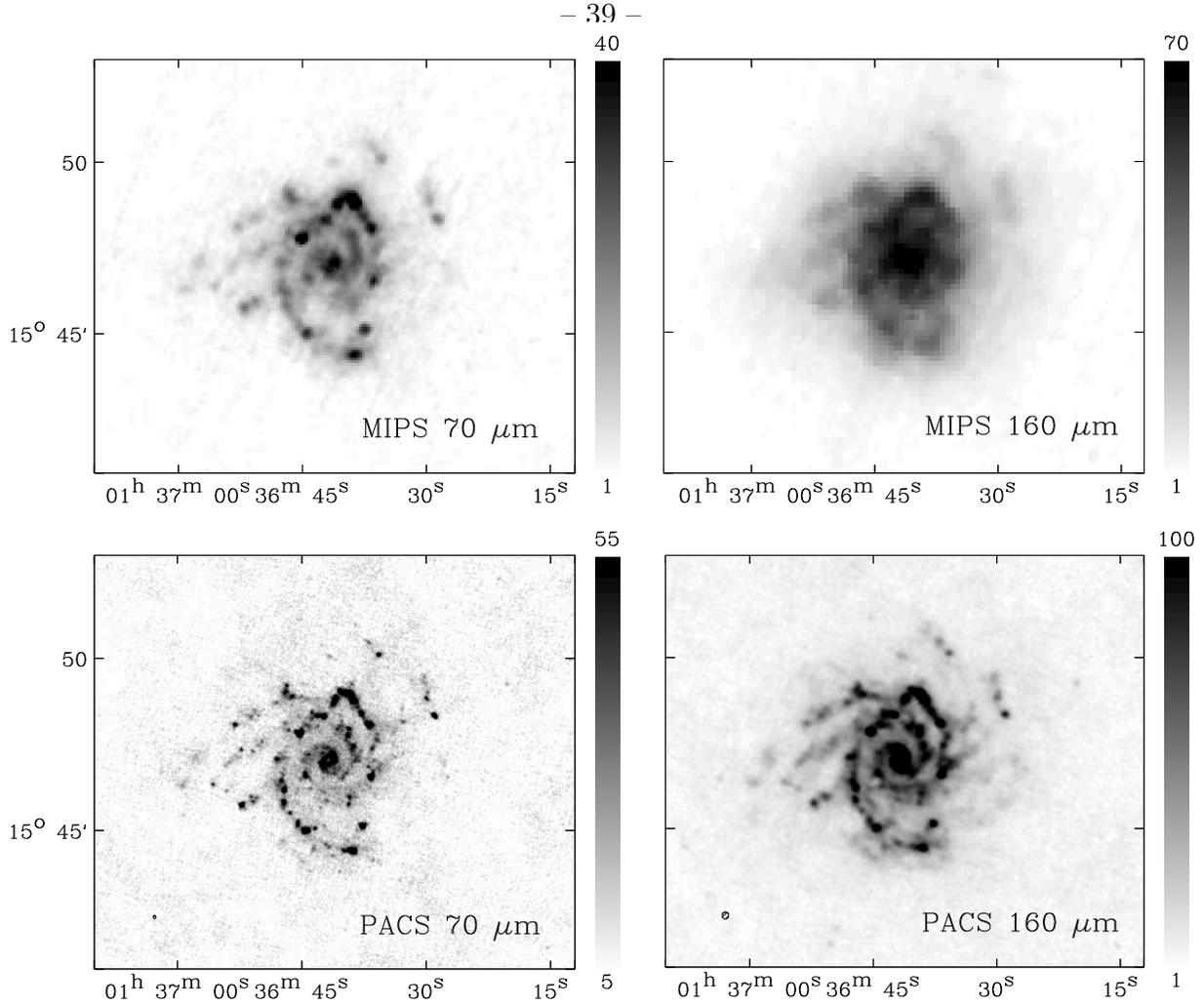}
%\vspace{-4cm}
\caption{ 
Top:  Far-infrared scan maps of the KINGFISH galaxy NGC\,628
at 70\,$\mu$m (left) and 160\,$\mu$m (right), as observed with
the {\it Spitzer} MIPS instrument as part of the SINGS project.  
Bottom:  Scan maps at the same wavelengths with the {\it Herschel}
PACS imager, and processed using the Scanamorphos mapping package.  
The superior spatial resolution of the PACS images
is readily apparent, though the MIPS maps should still be
more sensitive to faint extended emission. }
\end{figure}

\clearpage

\begin{figure}
\plotone{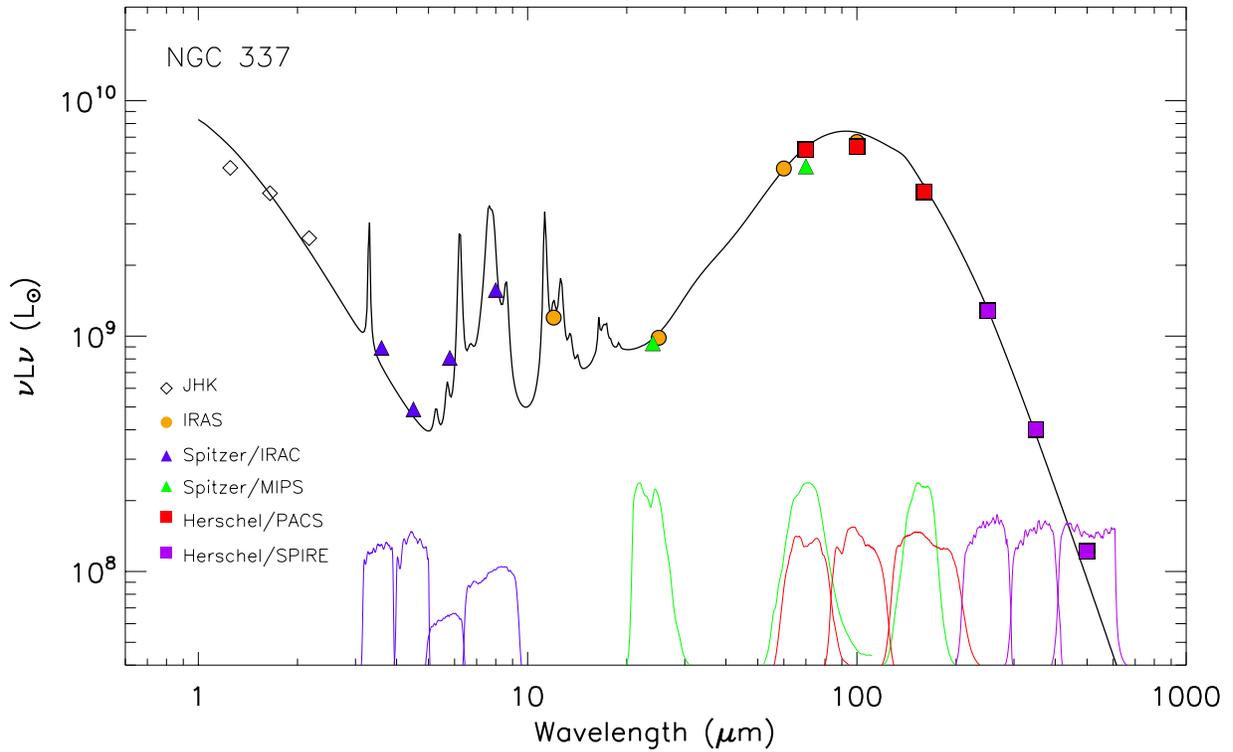}
\caption{Integrated spectral energy distribution (SED) for the
KINGFISH galaxy NGC\,337, based on combined measurements with
the {\it Spitzer} IRAC and MIPS instruments, and the {\it Herschel} PACS 
and SPIRE instruments.  The corresponding filter bandpasses
are shown at the bottom of the figure.  The dark line is a 
fit of a dust model following Draine et al. (2007).
Note the importance of the SPIRE fluxes in
the 250--500\,$\mu$m region for constraining the SED shape at
long wavelengths.}
\end{figure}

\clearpage

\begin{figure}
\epsscale{0.7}
\plotone{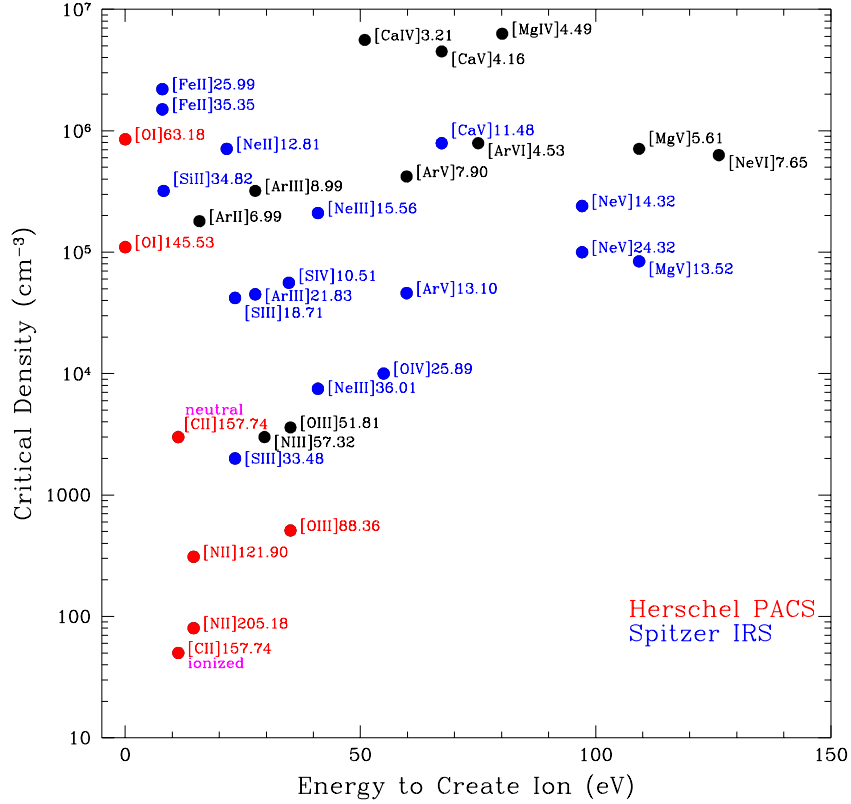}
\caption{Key diagnostic lines for the ISM,
in terms of ionization potential and critical density. 
Red points denote lines accessible within the spectral range of 
the {\it Herschel} PACS spectrometer, while blue points
show lines that lie within the spectral coverage of the
{\it Spitzer} IRS instrument.  Black points fall outside
of the sensitive range of either instrument but are shown
for completeness.  The two critical densities shown for the [CII]\,158\,$\mu$m
line apply to collisional de-excitations from electrons (ionized
regions) and from atoms (neutral regions).  
Note the unique range of conditions that are probed by 
the primary far-infrared cooling lines from {\it Herschel}.
All of the lines shown in red except for [OI]\,145.5\,$\mu$m 
are being observed as part of KINGFISH. }
\end{figure}

\clearpage
\begin{figure}
%\epsscale{0.80}
\plotone{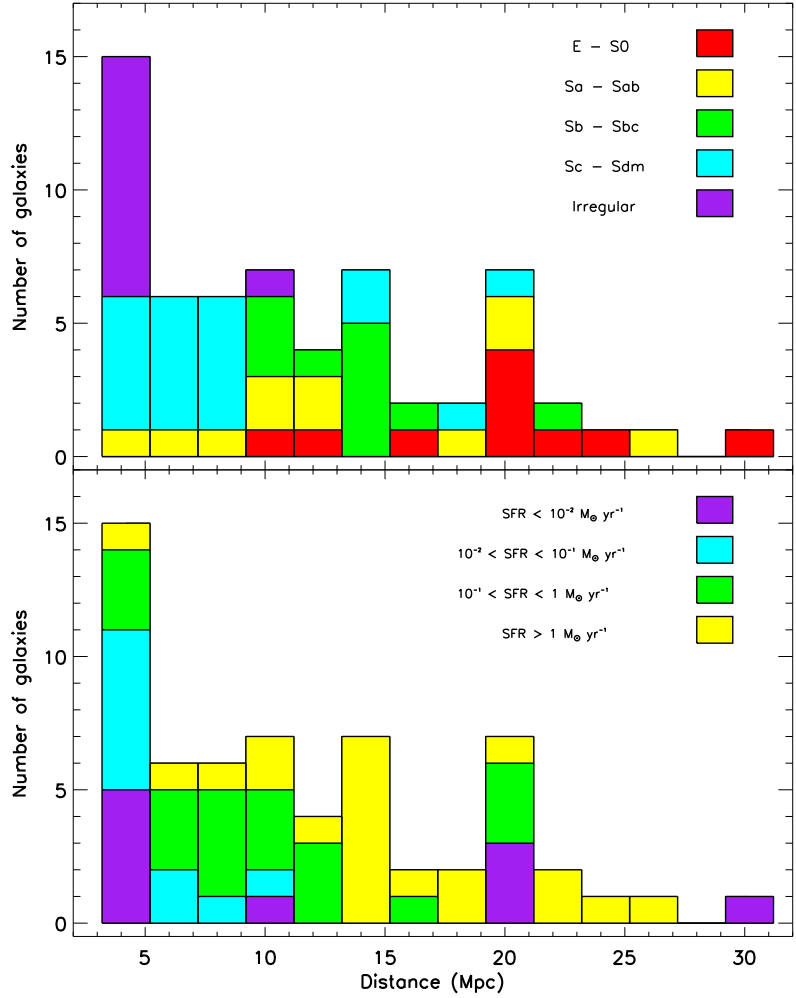} 
\caption{Above:  Distribution of KINGFISH galaxies by morphological
type and distance.  Below:  Distribution by star formation rate
(as derived from a combination of H$\alpha$ and 24\,$mu$m measurements) 
and distance.}
\end{figure}

\clearpage

\begin{figure}
\epsscale{1.0}
\plotone{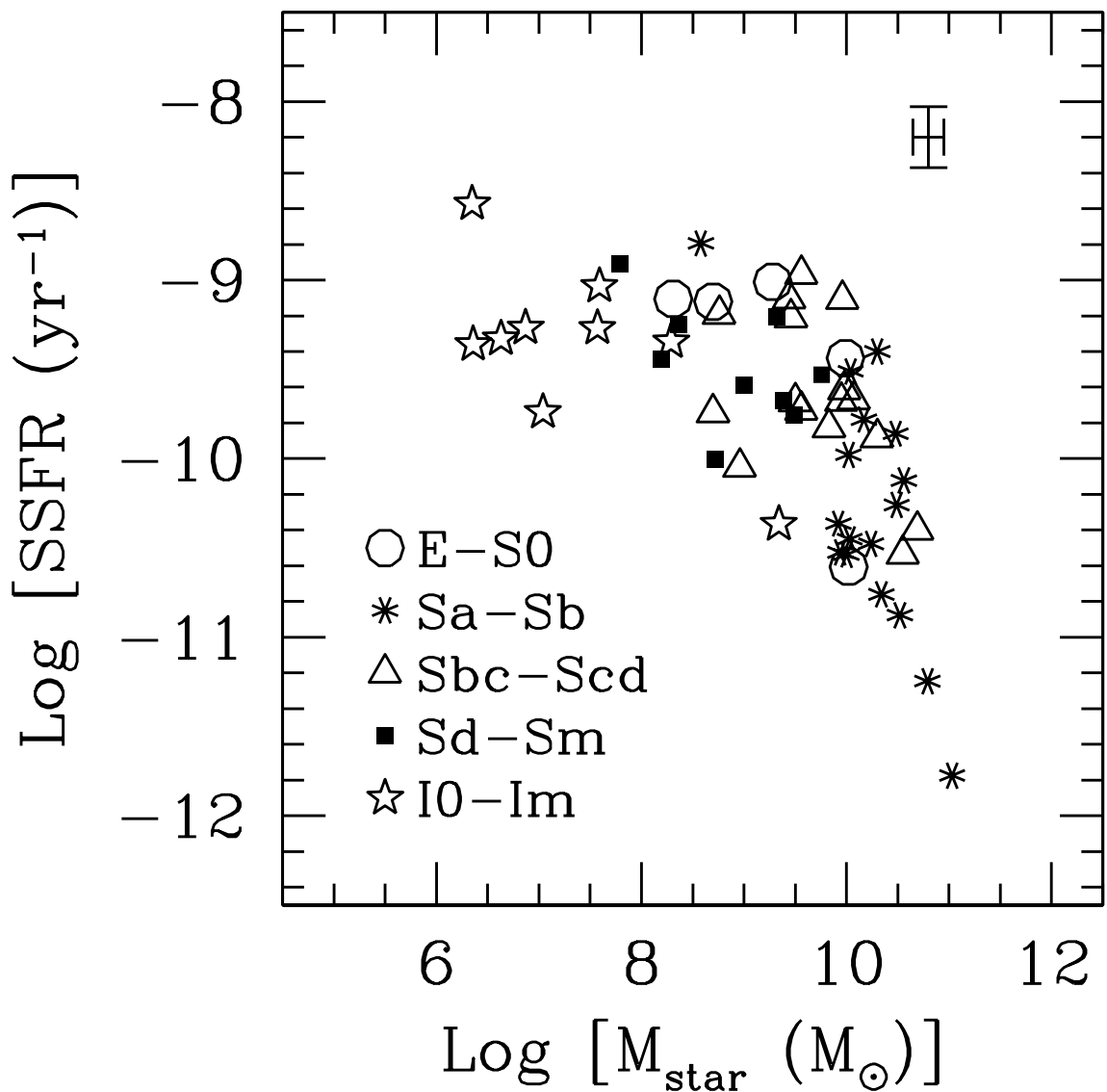}
\caption{Specific star formation rate for KINGFISH galaxies,
SSFR=SFR/M$_*$,
where the SFR and the stellar mass M$_*$ are from Table~1. Only galaxies
for which a SFR is available are shown on the plot. A representative
1~$\sigma$ error bar is shown in the top-right corner of the figure.
The error bar includes only random errors (Skibba et al. 2001).}
\end{figure}

\clearpage

\begin{figure}
%\epsscale{1.0}
%\plotone{f6.eps}
\includegraphics[scale=.40]{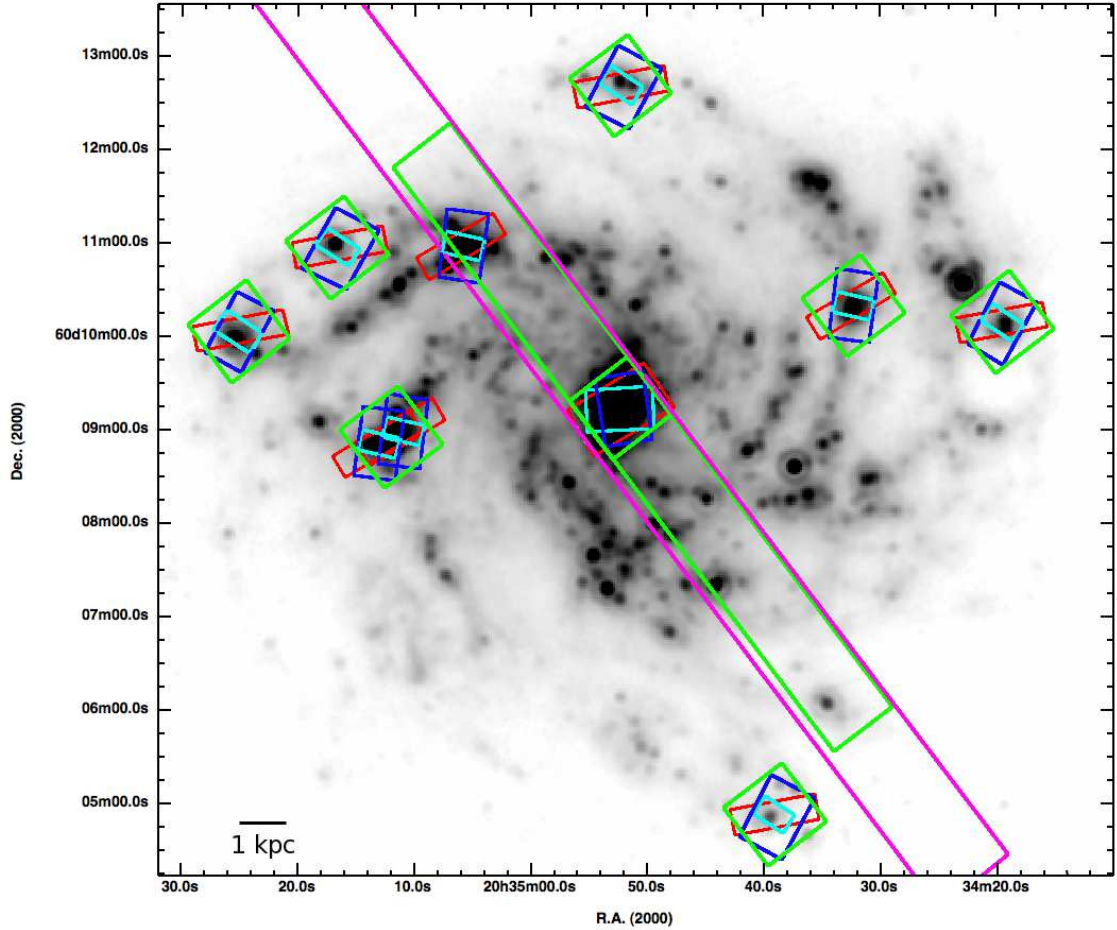}
\caption{{\it Spitzer} 8\,\micron\ image of NGC 6946,
with PACS spectroscopic line imaging regions superimposed
(green boxes).  The other color boxes show the footprints
of the corresponding {\it Spitzer} IRS observations from SINGS:
5--14\,$\mu$m low resolution spectral maps (red); 
14--38\,$\mu$m low resolution maps (magenta); and 10--37\,$\mu$m
high-resolution spectral maps (cyan and blue).}
\end{figure}

\clearpage

\begin{figure}
%\epsscale{1.0}
%\plotone{f7.eps}
\includegraphics[angle=0,scale=.60]{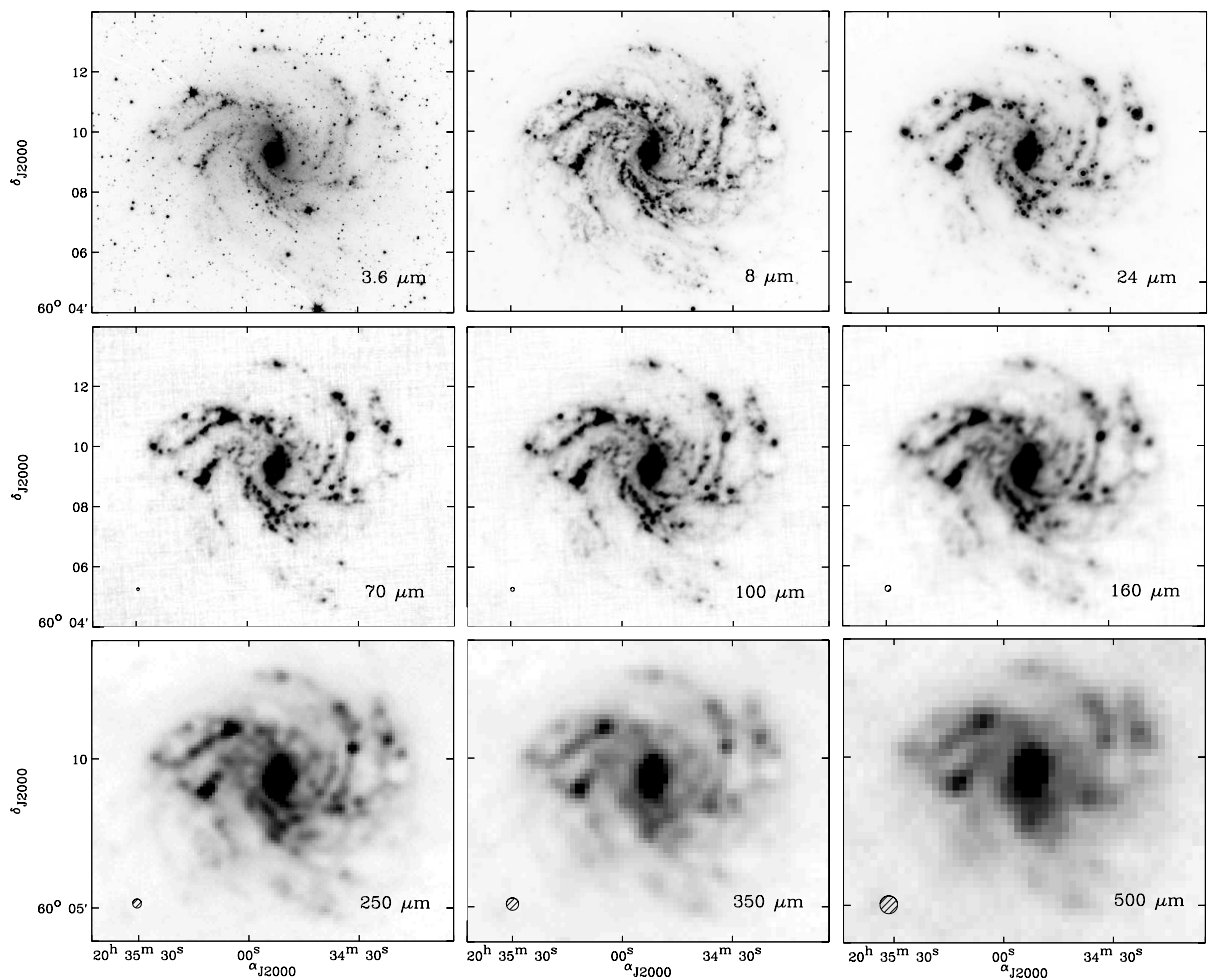}
\caption{A montage of infrared images of NGC\,6946 from
{\it Spitzer} (SINGS) and {\it Herschel} (KINGFISH).  FWHM beam sizes
for the respective {\it Herschel} bands are shown in the lower
left corner of each panel.   {\it Top panels}:  {\it Spitzer} IRAC
images at 3.6\,\micron\ and 8.0\,\micron, and MIPS
image at 24\,\micron.  The emission at these wavelengths
is dominated by stars, small PAH dust grains, and 
small dust grains heated by intense radiatino fields, respectively.
{\it Middle panels}:  {\it Herschel} PACS images at 70\,\micron, 
100\,\micron, and 160\,\micron, processed with the Scanamorphos
map making package.  Note the excellent 
spatial resolution despite the longer wavelengths, and
the progressive increase in contributions from diffuse
dust emission (``cirrus") with increasing wavelength.
{\it Bottom panels}:  {\it Herschel} SPIRE images at 250\,\micron,
350\,\micron, and 500\,\micron. These bands trace increasingly
cooler components of the main thermal dust emission, with
possible additional contributions from ``submillimeter 
excess" emission at the longest SPIRE wavelengths. {\it A figure
set with similar images for all 61 galaxies in the KINGFISH sample
can be found in the on-line version of this article. }}
\end{figure}

\clearpage

\begin{figure}
\epsscale{1.0}
\plotone{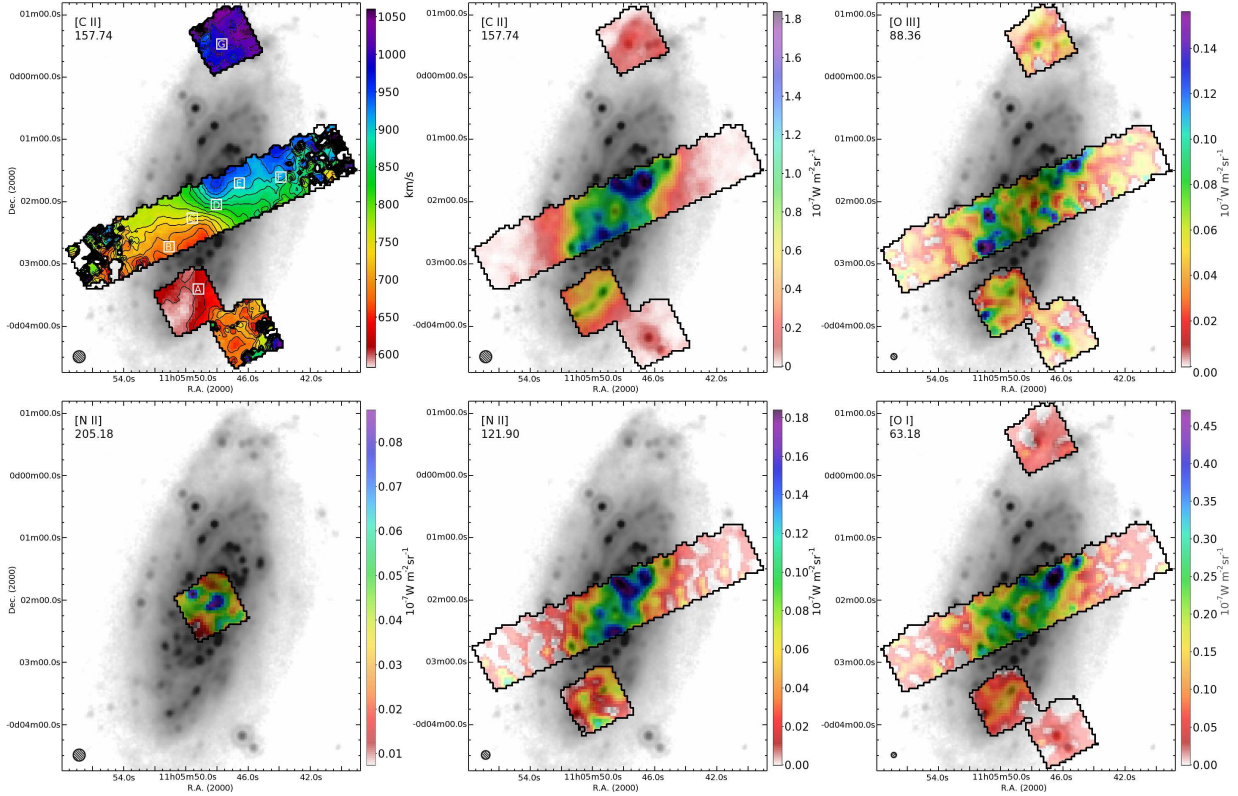}
\caption{Line maps of [CII], [OIII], the two [NII] lines, and [OI]
from KINGFISH observations of NGC\,3521.  The line maps are overlaid on
the SINGS MIPS 24 $\mu$m image, and clipped at an RMS level of  3$\sigma$.
The top left panel shows radial velocities as measured from the 
[CII] line (color and contours), while the other panels show
line flux, in units given by the respective color bars.  Note the
range of flux between different lines.  The white boxes in the top
left panel indicate the extraction apertures for the line profiles 
that are shown in Figure 9.}
\end{figure}

\clearpage

\begin{figure}
\epsscale{1.0}
\plotone{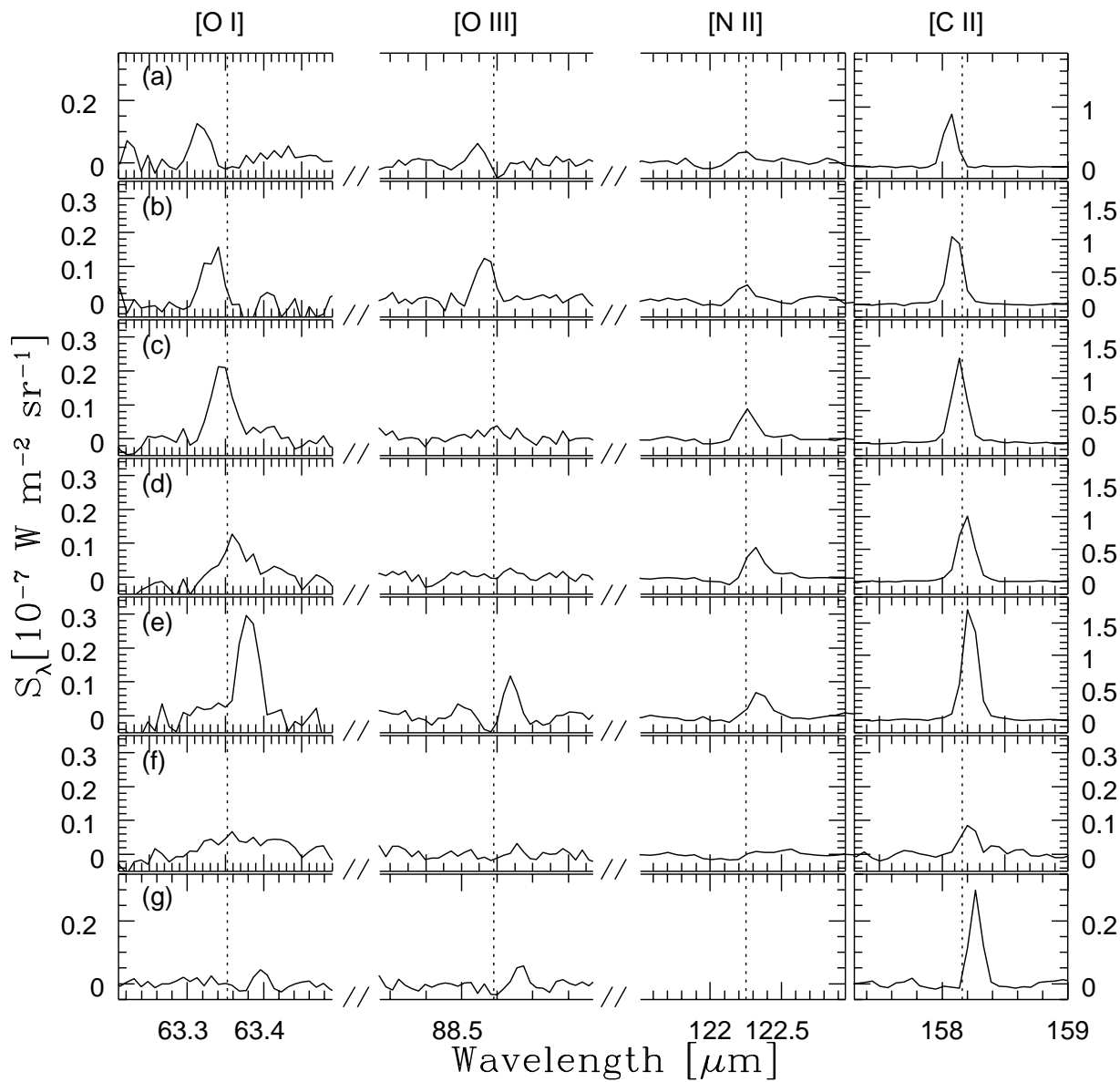}
\caption{Spectra of [OI], [OIII], [NII], and [CII]  are shown for
representative apertures (Figure 8) extracted from KINGFISH observations of NGC\,3521.
Apertures cover a range of surface brightness and environments, e.g.,
diffuse gas and HII regions.  The vertical lines show the expected
position of the lines at the systemic velocity of NGC\,3521; the offsets
are caused by the rotation of the galaxy. }
\end{figure}

\clearpage
\epsscale{1.0}
\begin{figure}
\plotone{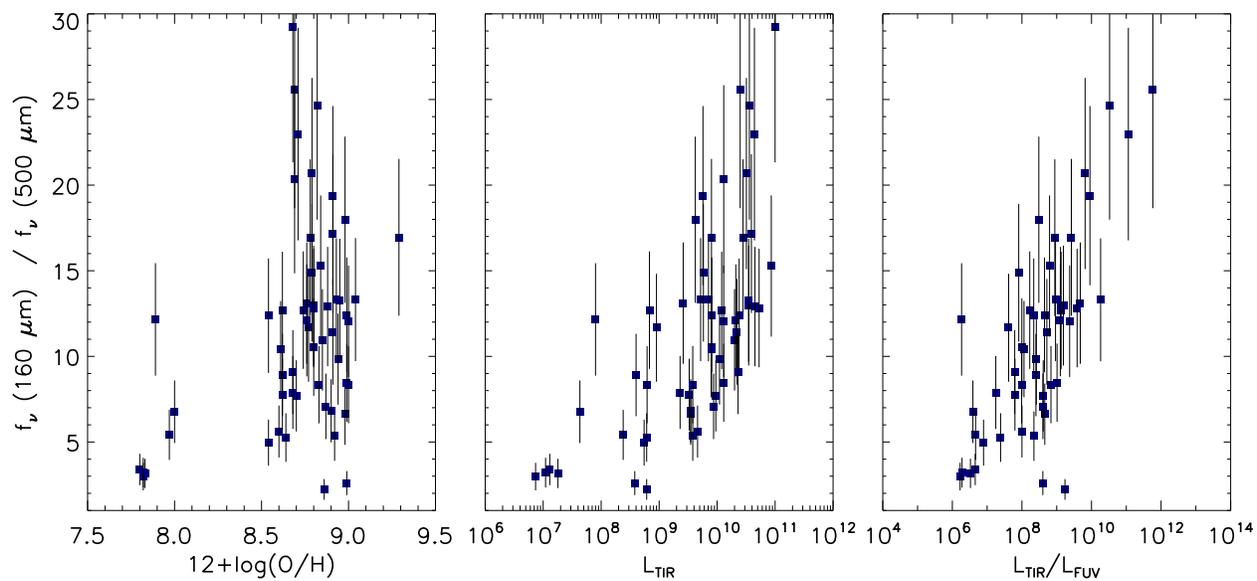}
\caption{Ratio of integrated 160\,$\mu$m flux density (MIPS) to
  500\,$\mu$m flux density (SPIRE) for the KINGFISH galaxies, 
  plotted as functions of mean disk metal abundance total
  infrared luminosity, and infrared to far-ultraviolet
  (153\,nm) flux ratio, an approximate measure of total dust obscuration.
  Note the strong trends, in sense of increased submillimeter
  emission and thus {\it colder} dust SEDs in the fainter,
  less dusty, and more metal-poor lower-mass galaxies.}
\end{figure}

\begin{deluxetable}{lrrrrrrrllllrrc}
\tablecolumns{15}
\rotate
\tabletypesize{\scriptsize}
\tablecaption{Properties of the KINGFISH Galaxies Sample.\label{tbl-1}}
\tablewidth{0pt}
\tablehead{
\colhead{Name\tablenotemark{a}} & \colhead{v$_H$\tablenotemark{a}} 
& \colhead{Morph.\tablenotemark{a}}  & \colhead{Sizes\tablenotemark{a}} 
& \colhead{Dist.\tablenotemark{b}} & \colhead{Method\tablenotemark{c}}
& \colhead{Ref\tablenotemark{d}}
& \colhead{Nuc. Type\tablenotemark{e}} & \multicolumn{2}{c}{12$+$log(O/H)\tablenotemark{f}} 
%& \colhead{12$+$log(O/H)\tablenotemark{f}} 
& \colhead{L$_{3.6}$\tablenotemark{g}} 
& \colhead{L$_{TIR}$\tablenotemark{h}} 
& \colhead{SFR\tablenotemark{i}} & \colhead{M$_*$\tablenotemark{j}} 
& \colhead{PACS/SPIRE\tablenotemark{k}} 
\\
\colhead{} & \colhead{(km~s$^{-1}$)} 
& \colhead{} & \colhead{($^{\prime}\times ^{\prime}$)} 
& \colhead{(Mpc)} & \colhead{} 
& \colhead{}  & \colhead{} & \colhead{(PT)} & \colhead{(KK)}
& \colhead{(L$_{\odot}$)} & \colhead{(L$_{\odot}$)} 
& \colhead{(M$_{\odot}$~yr$^{-1}$)} & \colhead{(M$_{\odot}$)}
& \colhead{}
\\
\colhead{(1)} & \colhead{(2)} & \colhead{(3)} & \colhead{(4)} & \colhead{(5)} & 
\colhead{(6)} & \colhead{(7)} & \colhead{(8)} & \colhead{(9)} & \colhead{(10)} &
\colhead{(11)} & \colhead{(12)} & \colhead{(13)} & \colhead{(14)} & \colhead{(15)} 
\\
}
\startdata
NGC0337 & 1650 & SBd    &  2.9$\times$1.8 & 19.3  & TF   &  1 & SF        & 8.18    & 8.84    & 9.4E8    & 1.2E10 & 1.30  &  9.32 &  B \\ 
NGC0584 & 1854 & E4     &  4.2$\times$2.3 & 20.8  & SBF  &  2 & ...       & 8.43(+) & 9.07(+) & 4.2E9    & 6.1E8  & ...   & 11.12 &  F  \\ 
NGC0628 &  657 & SAc    & 10.5$\times$9.5 &  7.2  & SNII &  3 & ...       & 8.35    & 9.02    & 1.2E9    & 8.0E9  & 0.68  &  9.56 &  B \\ 
NGC0855 &  610 & E      &  2.6$\times$1.0 &  9.73 & SBF  &  2 & SF        & 8.29    & 8.80    & 1.1E8    & 4.0E8  & ...   &  8.67 &  B \\ 
NGC0925 &  553 & SABd   & 10.5$\times$5.9 &  9.12 & Ceph &  4 & SF        & 8.25    & 8.79    & 6.7E8    & 4.6E9  & 0.54  &  9.49 & F  \\ 
NGC1097 & 1275 & SBb    &  9.3$\times$6.3 & 14.2  & TF   &  5 & AGN       & 8.47    & 9.09    & 6.5E9    & 4.5E10 & 4.17  & 10.48 & B  \\ 
NGC1266 & 2194 & SB0    &  1.5$\times$1.0 & 30.6  & TF   &  6 & AGN       & 8.29(+) & 8.89(+) & 1.3E9    & 2.5E10 & ...   & 10.14 &  B \\ 
NGC1291 &  839 & SBa    &  9.8$\times$8.1 & 10.4  & TF   &  6 & AGN       & 8.52(+) & 9.20(+) & 5.9E9    & 3.5E9  & 0.35  & 10.79 &  B \\ 
NGC1316 & 1760 & SAB0   & 12.0$\times$8.5 & 21.0  & SBF  &  7 & AGN       & 8.77(+) & 9.52(+) & 2.9E10   & 8.0E9  & ...   & 11.46 &  F \\ 
NGC1377 & 1792 & S0     &  1.8$\times$0.9 & 24.6  & TF   &  6 & ...       & 8.29(+) & 8.89(+) & 9.0E8    & 1.3E10 & 1.86  &  9.28 &  B \\ 
NGC1404 & 1947 & E1     &  3.3$\times$3.0 & 20.2  & SBF  &  7 & ...       & 8.54(+) & 9.21(+) & 7.8E9    & 3.8E8  & ...   & 10.88 &  F \\ 
IC342   &   31 & SABcd  & 21.4$\times$20. &  3.28 & Ceph &  8 & SF(*)     & 8.49(P) & ...     & 2.5E9(E) & 1.4E10 & 1.87  &  9.95 &  B \\ 
NGC1482 & 1655 & SA0    &  2.5$\times$1.4 & 22.6  & TF   &  6 & SF        & 8.11    & 8.95    & 2.8E9    & 4.4E10 & 3.57  &  9.99 &  B \\ 
NGC1512 &  896 & SBab   &  8.9$\times$5.6 & 11.6  & TF   &  1 & AGN       & 8.56    & 9.11    & 1.4E9    & 3.8E9  & 0.36  &  9.92 &  F \\ 
NGC2146 &  893 & Sbab   &  6.0$\times$3.4 & 17.2  & TF   &  9 & SF(*)     & 8.68(E) & ...     & 7.3E9(E) & 1.0E11 & 7.94  & 10.30 &  F \\ 
HoII    &  157 & Im     &  7.9$\times$6.3 &  3.05 & Ceph & 10 & ...       & 7.72    & 8.13    & 1.7E7    & 8.0E7  & 0.036 &  7.59 &  F  \\ 
DDO053  &   19 & Im     &  1.5$\times$1.3 &  3.61 & TRGB & 11 & ...       & 7.60    & 8.00    & 1.7E6    & 1.3E7  & 0.006 &  6.35 & B  \\ 
NGC2798 & 1726 & SBa    &  2.6$\times$1.0 & 25.8  & TF   &  6 & SF/AGN    & 8.34    & 9.04    & 1.9E9    & 3.6E10 & 3.38  & 10.04 & B  \\ 
NGC2841 &  638 & SAb    &  8.1$\times$3.5 & 14.1  & Ceph & 12 & AGN       & 8.54    & 9.21    & 6.6E9    & 1.3E10 & 2.45  & 10.17 &  B \\ 
NGC2915 &  468 & I0     &  1.9$\times$1.0 &  3.78 & TRGB & 13 & SF        & 7.94    & 8.28    & 2.0E7    & 4.3E7  & 0.020 &  7.57 &  F \\ 
HoI     &  143 & IABm   &  3.6$\times$3.0 &  3.9  & TRGB & 11 & ...       & 7.61    & 8.04    & 4.8E6    & 1.8E7  & 0.004 &  6.87 &  B \\ 
NGC2976 &    3 & SAc    &  5.9$\times$2.7 &  3.55 & TRGB & 11 & SF        & 8.36    & 8.98    & 1.4E8    & 9.0E8  & 0.082 &  8.96 & B  \\ 
NGC3049 & 1494 & SBab   &  2.2$\times$1.4 & 19.2  & TF   &  6 & SF        & 8.53    & 9.10    & 3.8E8    & 3.5E9  & 0.61  &  8.58 &  B \\ 
NGC3077 &   14 & I0pec  &  5.4$\times$4.5 &  3.83 & TRGB & 11 & SF(*)     & ...     & 8.9:(C) & 2.1E9(D) & 6.4E8  & 0.094 &  9.34 &  F \\ 
M81DwB  &  350 & Im     &  0.9$\times$0.6 &  3.6\tablenotemark{l}  & M81G &    & ...       & 7.84    & 8.19    & 1.7E6    & 6.8E6  & 0.001 &  6.36 & B  \\ 
NGC3190 & 1271 & SAap   &  4.4$\times$1.5 & 19.3  & TF   &  6 & AGN(*)    & 8.49(+) & 9.15(+) & 3.6E9    & 7.1E9  & 0.38  & 10.03 & B  \\ 
NGC3184 &  592 & SABcd  &  7.4$\times$6.9 & 11.7  & SNII & 14 & SF        & 8.51    & 9.15    & 2.0E9    & 1.1E10 & 0.66  &  9.50 &  B \\ 
NGC3198 &  663 & SBc    &  8.5$\times$3.3 & 14.1  & Ceph & 15 & SF        & 8.34    & 8.90    & 1.4E9    & 9.5E9  & 1.01  &  9.83 &  F \\ 
IC2574  &   57 & SABm   & 13.2$\times$5.4 &  3.79 & TRGB & 11 & SF(*)     & 7.85    & 8.24    & 5.6E7    & 2.4E8  & 0.057 &  8.20 &  F  \\ 
NGC3265 & 1421 & E      &  1.3$\times$1.0 & 19.6: & TF   &  6 & SF        & 8.27    & 8.99    & 2.8E8    & 2.6E9  & 0.38  &  8.70 &  B \\ 
NGC3351 &  778 & SBb    &  7.4$\times$5.0 &  9.33 & Ceph &  4 & SF        & 8.60    & 9.19    & 1.8E9    & 8.1E9  & 0.58  & 10.24 &  B \\ 
NGC3521 &  805 & SABbc  & 11.0$\times$5.1 & 11.2  & TF   &  5 & SF/AGN(*) & 8.39    & 9.01    & 6.7E9    & 3.5E10 & 1.95  & 10.69 & B  \\ 
NGC3621 &  727 & SAd    & 12.3$\times$7.1 &  6.55 & Ceph &  4 & AGN       & 8.27    & 8.80    & 1.1E9    & 7.9E9  & 0.51  &  9.38 & B  \\ 
NGC3627 &  727 & SABb   &  9.1$\times$4.2 &  9.38 & Ceph &  4 & AGN       & 8.34    & 8.99    & 4.3E9    & 2.8E10 & 1.70  & 10.49 & B  \\ 
NGC3773 &  987 & SA0    &  1.2$\times$1.0 & 12.4  & TF   &  6 & SF        & 8.43    & 8.92    & 8.8E7    & 6.8E8  & 0.16  &  8.31 &  B \\ 
NGC3938 &  809 & SAc    &  5.4$\times$4.9 & 17.9  & SNII & 16 & SF(*)     & 8.42(+) & 9.06(+) & 2.7E9    & 2.0E10 & 1.77  &  9.46 & B  \\ 
NGC4236 &    0 & SBdm   & 21.9$\times$7.2 &  4.45 & TRGB & 17 & SF(*)     & 8.17    & 8.74(+) & 1.3E8    & 5.5E8  & 0.13  &  8.36 & B  \\ 
NGC4254 & 2407 & SAc    &  5.4$\times$4.7 & 14.4  & SNII & 16 & SF/AGN    & 8.45    & 9.13    & 3.8E9    & 3.9E10 & 3.92  &  9.56 &  B \\ 
NGC4321 & 1571 & SABbc  &  7.4$\times$6.3 & 14.3  & Ceph &  4 & AGN       & 8.50    & 9.17    & 5.0E9    & 3.5E10 & 2.61  & 10.30 & B  \\ 
NGC4536 & 1808 & SABbc  &  7.6$\times$3.2 & 14.5  & Ceph &  4 & SF/AGN    & 8.21    & 9.00    & 2.2E9    & 2.1E10 & 2.17  &  9.44 &  F \\ 
NGC4559 &  816 & SABcd  & 10.7$\times$4.4 &  6.98 & TF   &  5 & SF        & 8.29    & 8.81    & 4.4E8    & 3.3E9  & 0.37  &  8.76 & F  \\ 
NGC4569 & -235 & SABab  &  9.5$\times$4.4 &  9.86 & TF   & 18 & AGN       & 8.58(+) & 9.26(+) & 1.9E9    & 5.2E9  & 0.29  & 10.00 &  F \\ 
NGC4579 & 1519 & SABb   &  5.9$\times$4.7 & 16.4  & TF   &  1 & AGN       & 8.54(+) & 9.22(+) & 6.1E9    & 1.3E10 & 1.10  & 10.02 & B  \\ 
NGC4594 & 1091 & SAa    &  8.7$\times$3.5 &  9.08 & SBF  & 19 & AGN       & 8.54(+) & 9.22(+) & 8.5E9    & 3.8E9  & 0.18  & 11.03 & F  \\ 
NGC4625 &  609 & SABmp  &  2.2$\times$1.9 &  9.3  & TF   &  6 & SF        & 8.35    & 9.05    & 1.1E8    & 6.2E8  & 0.052 &  8.72 & B  \\ 
NGC4631 &  606 & SBd    & 15.5$\times$2.7 &  7.62 & TRGB & 20 & SF(*)     & 8.12    & 8.75    & 1.9E9    & 2.4E10 & 1.70  &  9.76 & B  \\ 
NGC4725 & 1206 & SABab  & 10.7$\times$7.6 & 11.9  & Ceph &  4 & AGN       & 8.35    & 9.10    & 4.2E9    & 8.7E9  & 0.44  & 10.52 & F  \\ 
NGC4736 &  308 & SAab   & 11.2$\times$9.1 &  4.66 & TRGB & 21 & AGN(*)    & 8.31    & 9.01    & 2.0E9    & 5.8E9  & 0.38  & 10.34 & B  \\ 
DDO154  &  376 & IBm    &  3.0$\times$2.2 &  4.3  & BS   & 22 & ...       & 7.54    & 8.02    & 1.9E6    & 7.4E6  & 0.002 &  6.63 &  F  \\ 
NGC4826 &  408 & SAab   & 10.0$\times$5.4 &  5.27 & TRGB & 23 & AGN       & 8.54    & 9.20    & 1.8E9    & 4.2E9  & 0.26  &  9.94 &  F \\ 
DDO165  &   37 & Im     &  3.5$\times$1.9 &  4.57 & TRGB & 17 & ...       & 7.63    & 8.04    & 8.7E6    & 1.1E7  & 0.002 &  7.04 &   F \\ 
NGC5055 &  504 & SAbc   & 12.6$\times$7.2 &  7.94 & TF   &  5 & AGN       & 8.40    & 9.14    & 3.9E9    & 2.2E10 & 1.04  & 10.55 & B  \\ 
NGC5398 & 1216 & SBdm   &  2.8$\times$1.7 &  7.66 & TF   &  1 & ...       & 8.35    & 8.69    & 5.8E7    & 3.9E8  & 0.076 &  7.79 &  B \\ 
NGC5408 &  509 & IBm    &  1.6$\times$0.8 &  4.8  & TRGB & 24 & ...       & 7.81    & 8.23    & 3.1E7    & 1.9E8  & 0.088 &  8.29 & B  \\ 
NGC5457 &  241 & SABcd  & 28.8$\times$26. &  6.7  & Ceph &  4 & SF(*)     & 8.68(B) & ...     & 3.2E9(D) & 2.3E10 & 2.33  &  9.98 &  F  \\ 
NGC5474 &  273 & SAcd   &  4.8$\times$4.3 &  6.8  & BS   & 25 & SF(*)     & 8.31    & 8.83    & 1.2E8    & 6.1E8  & 0.091 &  8.70 & B  \\ 
NGC5713 & 1883 & SABbcp &  2.8$\times$2.5 & 21.4  & TF   & 26 & SF        & 8.24    & 9.03    & 2.4E9    & 3.2E10 & 2.52  & 10.07 & B  \\ 
NGC5866 &  692 & S0     &  4.7$\times$1.9 & 15.3  & SBF  &  2 & AGN       & 8.47(+) & 9.12(+) & 4.0E9    & 5.7E9  & 0.26  & 10.02 &  B \\ 
NGC6946 &   48 & SABcd  & 11.5$\times$9.8 &  6.8  & TRGB & 27 & SF        & 8.40    & 9.05    & 4.0E9    & 8.6E10 & 7.12  &  9.96 & B  \\ 
NGC7331 &  816 & SAb    & 10.5$\times$3.7 & 14.5  & Ceph &  4 & AGN       & 8.34    & 9.02    & 8.8E9    & 5.3E10 & 2.74  & 10.56 &  B \\ 
NGC7793 &  230 & SAd    &  9.3$\times$6.3 &  3.91 & TRGB & 28 & SF        & 8.31    & 8.88    & 3.1E8    & 2.3E9  & 0.26  &  9.00 & B  \\ 
%% Te$\times$t for table notes should follow after the \enddata but before
%% the \end{delu$\times$etable}. Make sure there is at least one \tablenotemark
%% in the table for each \tablenotete$\times$t.
\enddata

%\tablenotetext{a}{...}
%\tablenotetext{b}{...}
%\tablenotetext{c}{...}
%\tablenotetext{d}{...}
%\tablenotetext{e}{...}
%\tablenotetext{f}{...}
%\tablenotetext{g}{...}
%\tablenotetext{h}{...}
%\tablenotetext{i}{...}

\tablenotetext{a}{Galaxy name, Heliocentric recession velocity, morphological type, and sizes, as listed 
in NED, the NASA Extragalactic Database.}
\tablenotetext{b}{Redshift--independent distance in Mpc.}
\tablenotetext{c}{Methods employed to determine the redshift--independent distances. In order of decreasing 
preference: Cepheids (Ceph), Tip of the Red Giant Branch Stars (TRGB), Surface Brightness Fluctuations 
(SBF), Supernova Type II Plateau (SNII), flow--corrected Tully--Fisher relation (TF), bright stars (BS), and 
mean distance to the M81 group (M81G, 3.6 Mpc).}
\tablenotetext{d}{References to the distances: 1 -- Springob, C.M., Masters, K.L., Haynes, M.P., Giovanelli, R., \& Marinoni,
    C. 2009, ApJS, 182, 474;  2 -- Tonry, J.L., Dressler, A., Blakeslee, J.P., Ajhar, E.A., Fletcher, A.B., 
    et al. 2001, ApJ, 546, 681;  3 -- Van Dyk, S.D., Li, W., \& Filippenko, A.V. 2006, PASP, 118, 351;  4 -- Freedman, W.L., Madore, B.F., Gibson, B.K., Ferrarese, L., Kelson, D.D.,  et al. 2001, ApJ, 553, 47;  5 -- Tully, R.B., Rizzi, L., Shaya, E.J., Courtois, H.M., Makarov, D.I., \& 
   Jacobs, B.A. 2009, AJ, 138, 323;  6 -- K.L. Masters, 2005, private communication;  7 -- Blakeslee, J.P., Jordan, A., Mei, S., Cote, P., Ferrarese, L., Infante,   L. et al. 2009, ApJ, 694, 556;  8 -- Saha, A., Claver, J., \& Hoessel, J.G. 2002, AJ, 124, 839;  9 -- Tully, R.B. 1988, Nearby Galaxy Catalog; 10 -- Hoessel, J.G., Saha, A., Danielson, G.E. 1998, AJ, 115, 573;  11 -- Dalcanton, J.J., Williams, B.F., Seth, A.C., Dolphin, A., Holtzman, J.,  et al. 2009, ApJS, 183, 67;  12 -- Saha, A., Thim, F., Tamman, G.A., Reindl, B., \& Sandage, A. 2006, 
  ApJS, 165, 108;  13 -- Karachentsev, I.D., Makarov, D.I., Sharina, M.E., Dolphin, A.E., Grebel, 
     E.K., et al. 2003, A\&A 398, 479;  14 -- Jones, M.I., Hamuy, M., Lira, P., Maza, J., Clocchiatti, A., Phillips, 
        M. et al. 2009, ApJ, 696, 1176;  15 -- Kanbur, S.M., Ngeow, C., Nikolaev, S., Tanvir, N.R., \& Hendry, M.A. 
     2003, A\&A, 411, 361; 16 -- Poznanski, D., Nathaniel, B., Filippenko, A.V., Ganeshalingam, M., Li,  W., et al. 2009, ApJ, 694, 1067;  17 -- Karachensev, I.D., Dolphin, A.E., Geisler, D., Grebel, E.K., Guhatakurta, P., et al. 2002, A\&A , 383, 125;  18 -- Cortes, J.R., Kenney, J.D.P., \& Hardy, E. 2008, ApJ, 683, 78;  19 -- Jensen, J.B., Tonry, J.L., Barris, B.J., Thompson, R.I., Liu, M.C.,  Rieke, M.J., et al. 2003, ApJ, 583, 712;  20 -- Seth, A.C., Dalcanton, J.J., \& de Jong, R.S. 2005, AJ, 129, 1331;  21 -- Karachentsev, I.D., Sharina, M.E., Dolphin, A.E., Grebel, E.K., Geisler,
 D., et al. 2003, A\&A 398, 467;  22 -- Makarova, L., Karachentsev, I., Takalo, L.O., Heinaemaeki, P., \&  Valtonen, M. 1998, A\&AS, 128, 459;  23 -- Mould, J., \& Sakai, S. 2008, ApJ, 686, L75;  24 -- Karachentsev, I.D., Sharina, M.E., Dolphin, A.E., Grebel, E.K., Geisler,
 D., et al. 2002, A\&A, 385, 21;  25 -- Drozdovsky, I.O., \& Karachentsev, I.D. 2000, A\&AS, 142, 425;  26 -- Willick, J.A., Courteau, S., Faber, S.M., Burstein, D., Dekel, A., \&  Strauss, M.A. 1997, ApJS, 109, 333; 27 -- Karachentsev, I.D., Sharina, M.E., \& Huchmeier, W.K. 2000, A\&A, 362,  544;  28 -- Karachentsev, I.D., Grebel, E.K., Sharina, M.E., Dolphin, A.E., Geisler,  D., et al. 2003, A\&A, 404, 93.}
\tablenotetext{e}{Nuclear Type, based on optical spectroscopy: SF=Star Forming; 
     AGN=Non-Thermal Emission. From Table 5 of Moustakas, J., Kennicutt, 
     R.C., Tremonti, C.A., Dale, D.A., Smith, J.D.T., \& Calzetti, D. 2010, 
    ApJS, 190, 233; or (*) from Table 4 of Ho, L.C., Filippenko, A.V., \& 
    Sargent, W.L.W. 1997, ApJS, 112, 315.} 
\tablenotetext{f}{Characteristic oxygen abundances of the galaxies. The two columns, 
     (PT) and (KK), are the two oxygen abundances listed in Table 9 of Moustakas et al. (2010): the  
     PT value, in the left--hand--side column, is from the empirical calibration of 
     Pilyugin \& Thuan (2005); the KK value, in the right--hand--side column, is from 
     the theoretical calibration of Kobulnicky \& Kewley (2004). 
     (+) Oxygen abundance from the Luminosity-Metallicity relation. For the few galaxies 
     without oxygen abundances in Moustakas et al. only one metallicity value is 
     reported, from: (B) Bresolin, F., 
     Garnett, D. R., \& Kennicutt, R. C. 2004, ApJ, 615, 228. (C) Calzetti, 
     D., Harris, J., Gallagher, J.S., Smith, D.A., Conselice, C.J., 
     Homeier, N., \& Kewley, L. 2004, AJ, 127, 1405. (E) Engelbracth, C.W., 
     Rieke, G.H., Gordon, K.D., Smith, J.D.T., Werner, M.W., Moustakas, J., 
     Willmer, C.N.A., \& Vanzi, L. 2008, ApJ, 678, 804. (P) Pilyugin, L.S., 
     Vilchez, J.M., \& Contini, T. 2004, A\&A, 425, 849.}
\tablenotetext{g}{Integrated luminosity at 3.6~$\mu$m, in units of solar luminosity, expressed 
    as $\nu$l($\nu$), and derived from the fluxes listed in Dale, D.A., et al. 
    (2007), ApJ, 655, 863, and the distances in column (5). 
    (D) Fluxes for NGC3077 and NGC5457 from Dale, D.A., et al. 2009, ApJ, 
    703, 517. (E) Fluxes for IC342 and NGC2146 from Engelbracht, C.W., et al. 
    2008, ApJ, 678, 804. }
\tablenotetext{h}{Total infrared luminosity TIR in the 3--1100 $\mu$m range, in solar 
    luminosity units. Fluxes are from Spitzer photometry and converted to TIR 
    using equation (4) of Dale \& Helou (2002). Fluxes are either from Dale 
    et al. (2007) or from Dale et al. (2009). }
\tablenotetext{i}{Star Formation Rate (M$_{\odot}$~yr$^{-1}$), calculated from the combination 
    of H$\alpha$ and 24 $\mu$m luminosity given in Kennicutt et al. (2009). 
    Measurements as listed in Calzetti et al (2010). For NGC5457, the H$\alpha$ 
    luminosity is from Kennicutt et al. (2008) and the 24 micron luminosity 
    from Dale et al. (2009).}
\tablenotetext{j}{Stellar masses (M$_{\odot}$) obtained from the multi-color method described in 
   Zibetti et al. (2009), and listed in Skibba et al. (2011). The masses listed in Skibba et al. 
   are rescaled to our updated distance values.}
\tablenotetext{k}{PACS and SPIRE observation strategy for each galaxy. The galaxies in the KINGFISH sample 
   have been divided into a bright (B) and faint (F) bin, depending on their Spitzer/MIPS 160~$\mu$m surface brightness. 
   The bright bin has median surface brightness $\sim$3~MJy~sr$^{-1}$, while the faint bin has $\sim$1~MJy~sr$^{-1}$. 
   See text for more details.}
 \tablenotetext{l}{Although the mean distance of the M81 Group is used for this galaxy, an unpublished TRGB distance places M81DwB beyond the 
  Group.}
\end{deluxetable}

\begin{deluxetable}{lrccccc}
\tablecolumns{7}
%\rotate
%\tabletypesize{\scriptsize}
\tablecaption{PACS and SPIRE Imaging Observations}
\tablewidth{0pt}
\tablehead{
\colhead{Instrument} & \colhead{Band} 
& \colhead{PSF FWHM\tablenotemark{a}}  & \colhead{Pixel Size\tablenotemark{b}} 
& \colhead{} & \colhead{Sensitivity\tablenotemark{c}}  & \colhead{}  
\\
\colhead{} & \colhead{}
& \colhead{}  & \colhead{}
& \colhead{Bright} & \colhead{Faint}  & \colhead{Parallel}
\\
\colhead{} & \colhead{} 
& \colhead{arcsec} & \colhead{arcsec} 
& \colhead{MJy\,sr$^{-1}$} & \colhead{MJy\,sr$^{-1}$} & \colhead{MJy\,sr$^{-1}$}
}
\startdata
PACS &  70\,$\mu$m &  5.76 $\times$ 5.46 & 1.40 & 7.1  & 5.0  & 11.7 \\
PACS &  100\,$\mu$m &  6.89 $\times$ 6.69 & 1.70 & 7.1  & 5.0  & 12.3  \\
PACS &  160\,$\mu$m &  12.13 $\times$ 10.65 & 2.85 & 3.1  & 2.2 & 5.3  \\
SPIRE &  250\,$\mu$m &  18.3 $\times$ 17.0 & 6.0 & 1.0  & 0.7  & 2.5  \\ 
SPIRE &  350\,$\mu$m &  24.7 $\times$ 23.2 & 10.0 & 0.6  & 0.4  & 1.3  \\
SPIRE &  500\,$\mu$m &  37.0 $\times$ 33.4 & 14.0 & 0.3  & 0.2  & 0.6  \\
\enddata

\tablenotetext{a}{Major and minor axis PSF, as taken from PACS and SPIRE Observers' Manuals.  Values correspond to standard KINGFISH scan maps; the PSFs in Parallel mode (IC\,342 are slightly degraded, see Observer's Manuals.}
\tablenotetext{b}{Pixel sizes for processed data, see \S 4.}
\tablenotetext{c}{Anticipated 1-$\sigma$ sensitivities per pixel, estimated 
using HSPOT Version 5.  Updated sensitivities will be provided with data products upon delivery.}

\end{deluxetable}

\end{document}